\documentclass[prb,
superscriptaddress,showpacs,amsmath,amssymb]{revtex4}

\begin{document}

\author{G.E.~Volovik}
\affiliation{Low Temperature Laboratory, Aalto University,  P.O. Box 15100, FI-00076 Aalto, Finland}
\affiliation{Landau Institute for Theoretical Physics, acad. Semyonov av., 1a, 142432,
Chernogolovka, Russia}

\title{Thermodynamics and decay of de Sitter vacuum}

\date{\today}

\begin{abstract}
We discuss the consequences of the unique symmetry of the de Sitter spacetime. This symmetry leads to the specific thermodynamic properties of the de Sitter vacuum, which produces the thermal bath for matter. The de Sitter spacetime is invariant under the modified translations, ${\bf r}\rightarrow {\bf r} -e^{Ht}{\bf a}$, where $H$ is the Hubble parameter. For $H\rightarrow 0$, this symmetry corresponds to the conventional invariance of Minkowski spacetime under translations ${\bf r}\rightarrow {\bf r} -{\bf a}$. Due to this symmetry, all the comoving observers at any point of the de Sitter space perceive the de Sitter environment as the thermal bath with temperature $T=H/\pi$, which is twice larger than the Gibbons-Hawking temperature of the cosmological horizon. This temperature does not violate the de Sitter symmetry, and thus does not require the preferred reference frame, as  distinct from thermal state of matter, which violates the de Sitter symmetry. This leads to the heat exchange between gravity and matter, and to instability of de Sitter state towards the creation of matter, its further heating, and finally to the decay of the de Sitter state. The  temperature $T=H/\pi$ determines different processes in the de Sitter environment, which are not possible in the Minkowski vacuum, such as the process of ionization of an atom in the de Sitter environment. This temperature also determines the local entropy of the de Sitter vacuum state, and this allows us to calculate the total entropy inside the cosmological horizon. The result reproduces the  Gibbons-Hawking area law, which is related to the cosmological horizon, $S_{\rm hor}=4\pi KA$, where $K=1/(16\pi G)$. This supports the holographic properties of the cosmological event horizon. We extend the consideration of the local thermodynamics of the de Sitter state using the $f({\cal R})$ gravity. In this thermodynamics, the Ricci scalar curvature  ${\cal R}$ and the effective gravitational coupling $K$ are thermodynamically conjugate variables. The holographic connection between the bulk entropy of the Hubble volume and the surface entropy of the cosmological horizon remains the same, but with the gravitational coupling  $K=df/d{\cal R}$. Such connection takes place only in the $3+1$ spacetime, where there is the special symmetry due to which the variables $K$ and ${\cal R}$ have the same  dimensionality. We also consider the lessons from the de Sitter symmetry for the thermodynamics of black and white  holes.
\end{abstract}

\maketitle

\tableofcontents

\section{Introduction}

We consider the local thermodynamics of the de Sitter stage of the expansion of the Universe. The term ``local''  means that we consider the de Sitter vacuum as the thermal state, which is characterized by the local temperature.
This consideration is based on observation, that matter immersed in the de Sitter vacuum feels this vacuum as the heat bath with the local temperature $T=H/\pi$, where $H$ is the Hubble parameter. This temperature has no relation to the cosmological horizon, and to the Hawking radiation from the cosmological horizon. However,  it is exactly twice the Gibbons-Hawking temperature, $T_{\rm GH}=H/2\pi$. The reason for such relation is the symmetry of the de Sitter space-time with respect to the combined translations. In the Minkowski vacuum, this symmetry becomes the conventional invariance of under translations.

 The existence of the local temperature suggests the existence of the other local thermodynamic variables, which participate in the local thermodynamics of the de Sitter state. In addition to the the local entropy density $s$ and local vacuum energy density $\epsilon$, there are also the local thermodynamic variables related to the gravitational degrees of freedom. 

Although the local temperature is twice larger than the Gibbons-Hawking temperature assigned to the cosmological horizon, there is the certain connection between the local entropy $s$ and the global entropy usually assigned to the event horizon. It appears that the total entropy of the de Sitter state in the volume $V_H$ bounded by the cosmological horizon coincides with the Gibbons-Hawking entropy, which is proportional to the area $A$ of the horizon, $sV_H=A/4G$.  Such holographic bulk-surface correspondence takes place only in the $3+1$ spacetime. Although the peculiarity of the  $3+1$ spacetime may be related to the special symmetry, which connects the local thermodynamic variables,
 the origin of the holography is still not very clear. That is why we extended the consideration of the local thermodynamics to the $f({\cal R})$ gravity, and checked the bulk-surface correspondence in this modified version of general relativity.
 
 The $f({\cal R})$ gravity in terms of the Ricci scalar ${\cal R}$ is one of the simplest geometrical models,  which describes the dark energy and de Sitter expansion of the Universe.\cite{Starobinsky1980,Starobinsky2007,Felice2010,Clifton2012,Nojiri2010wj,Nojiri2017ncd,Odintsov2023b} It was used to construct an inflationary model of the early Universe -- the Starobinsky inflation, which is controlled by the ${\cal R}^2$ contribution to the effective action. This class of models, $f({\cal R}) \propto {\cal R} - {\cal R}^2/M^2$, was also reproduced in the so-called $q$-theory,\cite{KlinkhamerVolovik2008d,KlinkhamerVolovik2008c}
where $q$ is the 4-form field introduced by Hawking\cite{Hawking1984} for the phenomenological description of the physics of the deep (ultraviolet) vacuum (here the sign convention for ${\cal R}$ is opposite to that in Ref. \cite{Starobinsky2007}). The Starobinsky model is in good agreement
with the observations. However, despite the observational success, the theory of Starobinsky inflation is still phenomenological. Due to a rather small mass scale $M$ compared with the Planck scale it is difficult to embed the model into a UV complete theory.\cite{Chaichian2023,Brinkmann2023,Lust2023,Elizalde:2017mrn,Odintsov:2017hbk,Myrzakulov:2014hca,Bamba:2014jia,Sebastiani:2013eqa,Ketov2024} But we used this model only for the generalization of the de Sitter thermodynamics and for consideration of the validity of the holographic principle.
 
The $f({\cal R})$ theory demonstrates that the effective gravitational coupling $K$ (it is the inverse Newton constant, $K=1/16\pi G$) and the scalar curvature  ${\cal R}$ are connected by equation $K=df/d{\cal R}$. This suggests that $K$ and ${\cal R}$ are the thermodynamically conjugate variables.\cite{Volovik2023a,Volovik2023b}    This pair of the non-extensive gravitational variables is similar to the pair of the electrodynamic variables, electric field ${\bf E}$ and electric induction ${\bf D}$, which participate in the thermodynamics of dielectrics. They are also similar to the pair of magnetic thermodynamic variables, magnetic induction ${\bf B}$ and magnetic field ${\bf H}$. 

In the de Sitter spacetime, the local temperature does no depend on the theories of general relativity and thus has the same value $T=H/\pi$. Using the local thermodynamics with this temperature, we obtained the general result for the total entropy of the Hubble volume, $S_{\rm bulk}=sV_H=4\pi KA=A/4G=S_{\rm hor}$, where  $K=df/d{\cal R}$ is the effective gravitational coupling and ${\cal R}=-12H^2$. This supports the holographic bulk-surface correspondence in the $3+1$ spacetime. 

Note that the $3+1$ spacetime has special symmetry, which is absent in the other dimensions. The thermodynamically conjugate variables have the same dimensionality: all of them have dimensionality of the square of the inverse length: $[K]=[{\cal R}]=[1/l^2]$, $[{\bf B}]=[{\bf H}]=[1/l^2]$ and $[{\bf D}]=[{\bf E}]=[1/l^2]$. It looks that this symmetry is important for the validity of the holographic correspondence.

Since the de Sitter state has thermal behaviour, it serves as the thermal bath for matter. The matter violates the de Sitter symmetry, as a result the energy exchange between vacuum and matter leads to the decay of the de Sitter state.

The plan of the paper is the following. 

In Sec. \ref{HeatBathSec} we show that the de Sitter vacuum serves as the thermal bath for matter immersed in the de Sitter environment. Our approach differs from the traditional consideration of the vacua of the quantum fields in the de Sitter spacetime, which uses the Euclidean action method.
An example of the influence of the de Sitter vacuum on immersed matter is an atom in the de Sitter environment (Sec. \ref{AtomInDS}). As distinct from the atom in the flat space, the atom in the de Sitter vacuum has a certain probability of ionization. The rate of ionization is similar to the rate of ionization of an atom in the flat space-time in the presence of the thermal bath with temperature $T=H/\pi$.\cite{Volovik2009a,Volovik2009,Volovik2023c,Volovik2024c,Maxfield2022} 
In Sec. \ref{DecayComposite} it is shown that the same temperature determines the other activation processes, which are energetically forbidden in the Minkowski spacetime, but are allowed in the de Sitter background. Examples are the splitting of the heavy particle with mass $m$ to two particles with masses $m_1+m_2 >m$,\cite{Maldacena2015,Reece2023} and radiation of electron-positron pairs by the electron at rest. 
In Sec. \ref{ConnectionWithHawking} it is shown that the local temperature $T$ also determines the Gibbons-Hawking temperature of radiation from the cosmological horizon $T_{\rm GH}=T/2$ without using the Euclidean action.

Sec. \ref{dSthermodynamics} is devoted to the local thermodynamics of the de Sitter state, which is determined by the local temperature $T=H/\pi$. The local temperature leads to the local entropy of the de Sitter thermal bath (Sec. \ref{FromTtoEntropy}), which integrated over the Hubble volume reproduces the Gibbons-Hawking entropy of the cosmological horizon (Sec. \ref{HubbleVolume}).

Sec. \ref{HeatTransferSec} describes the attempt to obtain the de Sitter thermal states from the general principles of thermodynamics of the many-body systems. We consider the multi-metric gravity, which can be viewed as an ensemble of the sub-Universes, each being described by its own metric $g_{\mu\nu(n)}$ (or tetrads $e^a_{\mu(n)}$), by its own gravitational coupling  $K_n$ and cosmological constant $\Lambda_n$. The heat exchange between the sub-Universes leads to their thermalization -- the formation of the Universe in which the sub-Universes have the common Hubble parameter and thus they have the common temperature.

Sec. \ref{dSthermodynamicsSec} is devoted to thermodynamics of de Sitter in the $f({\cal R})$ gravity. The $f({\cal R})$ gravity contains the pair of the thermodynamically conjugate variables, $K$ and ${\cal R}$ (Sec. \ref{VariablesSec}). These variables together with the local temperature and local entropy provide the generalization of Gibbs-Duhem relation for the de Sitter state (Sec. \ref{GibbsExtensionSec}).  The confirmation of the holographic result for the total entropy of the Hubble volume in the $f({\cal R})$ gravity is obtained in Sec. \ref{EntropyKSec}: $S_{\rm bulk}=sV_H=4\pi KA=S_{\rm hor}$, The quadratic gravity and its symmetry are discussed in Sec. \ref{QuadraticSec}.

The Sec. \ref{decay} is devoted to the de Sitter decay due to the thermalization of matter by the de Sitter heat bath, and by the thermal fluctuations of the de Sitter state. These two mechanisms lead to different power laws of the decay, which may correspond to two different epochs. 

Sections \ref{BHsec} and \ref{macroTunnel} demonstrate how the local entropy of the de Sitter state allows us to consider the thermodynamics of the Schwarzschild black hole.
The starting point of our consideration in Sec. \ref{GravastarSec} is that the black hole  can be obtained from the relaxation of the gravastar object -- the black hole, which has the de Sitter core. It is important that the de Sitter interior of the gravastar is represented by the contracting de Sitter state. 
The contracting de Sitter in Sec. \ref{ContractingDSSec} has negative Hubble parameter $H<0$, and thus the negative temperature $T=H/\pi<0$ and the negative entropy. The contracting and expanding de Sitter states can be considered as two phases obtained from the symmetric state of the Minkowski vacuum by the spontaneous breaking of the time reversal symmetry. These phases transform to each other under the time reversal, $t\rightarrow -t$. In this sense, the Hubble parameter $H$ can be considered as the order parameter of the symmetry breaking phase transition.

Since in the considered gravastar object, the cosmological horizon and the black hole horizon cancel each other, the gravastar has zero entropy. In the process of  relaxation of the gravastar to the black hole, the de Sitter core with its negative entropy shrinks. This results in the Hawking-Bekenstein entropy of the black hole horizon (Sec. \ref{BHentropySec}). The negative entropy of the white hole is considered in Sec. \ref{antigravastarSec}. The heat exchange between black holes in the multi-metric gravity is discussed in Sec. \ref{ExchangeBetweenBHs}.
Section \ref{macroTunnel} provides the alternative derivation of the thermodynamics of black and white holes using the macroscopic quantum tunneling.

Finally the conclusion is in Sec. \ref{ConclusionSec}.

\section{de Sitter state as heat bath for matter}
\label{HeatBathSec}

\subsection{Atom in de Sitter environment as thermometer}
\label{AtomInDS}

We consider the de Sitter thermodynamics using the Painlev\'e-Gullstrand (PG) form,\cite{Painleve,Gullstrand} where the metric is
 \begin{equation}
ds^2= - dt^2 +   (d{\bf r} - {\bf v}({\bf r})dt)^2\,.
\label{PG1}
\end{equation}
Here ${\bf v}({\bf r})$ is the shift velocity, which in condensed matter plays the role of the velocity of the inviscid superfluid component of the liquid (the velocity of the "superfluid quantum vacuum" \cite{Volovik2003}).
In the de Sitter expansion the velocity of the "vacuum" is ${\bf v}({\bf r})=H{\bf r}$ and the metric is
 \begin{equation}
ds^2= - dt^2 +   (dr - Hr dt)^2+r^2 d\Omega^2\,.
\label{PG}
\end{equation}
 
The PG metric is stationary, i.e. does not depend on time, and it does not have the unphysical singularity at the cosmological horizon. That is why it is appropriate for consideration of the local thermodynamics both inside and outside the horizon. It also allows us to consider two different phases of the vacuum with broken time reversal symmetry. These are the expanding de Sitter Universe with $H>0$ and the contracting de Sitter Universe with $H<0$. These two phases transform to each other under the time reversal, $t\rightarrow -t$. In this sense, the Hubble parameter $H$ can be considered as the order parameter of the symmetry breaking phase transition from the symmetric state of the Minkowski vacuum.

Now let us consider an atom  at the origin, $r = 0$. The atom is the external object in the de Sitter spacetime, which violates the de Sitter symmetry. It plays the role of the detector (or the role of the static observer) in this spacetime. The electron bounded to an atom may absorb the energy from the gravitational field of the de Sitter background and escape from the electric potential barrier.  If the ionization potential is much smaller than the electron mass but is much larger than the Hubble parameter, $H\ll \epsilon_0 \ll m$, one can use
the nonrelativistic quantum mechanics to estimate the tunneling rate through the barrier. 

Let us consider an electron on the $n$-th level in the hydrogen atom. Under the de Sitter gravitational field this electron can escape from the atom with the conservation of energy, which in the classical limit is given by the classical equation: 
\begin{eqnarray}
\frac{p_r^2}{2m} +p_rv(r) = -E_n \,\,,\,\, E_n= \frac{me^4}{2\hbar^2}\frac{1}{n^2}\,.
\label{Classical1}
\end{eqnarray}
 Here $p_r$ is the momentum of electron in the radial direction, $v(r)=Hr$ and $p_r v(r)$ is the Doppler shift, which allows for electron to reach the negative energy $-E_n$ when it escapes from the atom.
 The corresponding radial trajectory $p_r(r)$ for the escape of an electron from the atom is:
\begin{eqnarray}
p_r(r)= -mv(r) + \sqrt{m^2v^2(r) -2m E_n}\,.
\label{ElectronTrajectory}
\end{eqnarray}
The integral of $p_r(r)$ over the classically forbidden region, $0 < r < r_n=\sqrt{2E_n/mH^2}$, gives the  ionization rate in the semiclassical WKB approximpation:
\begin{eqnarray}
w\sim \exp{ \left(-2\,{\rm Im}\, S\right)}=
\nonumber
\\
= \exp{ \left(-2\int_0^{r_n} dr \sqrt{2m E_n-m^2H^2r^2}\right)} =\exp\left(-\frac{\pi E_n}{H} \right)=\exp\left(-\frac{E_n}{T} \right)\,.
\label{IonizationRate}
\end{eqnarray}
This ionization rate is equivalent to the rate of ionization in the flat Minkowski spacetime in the presence of the heat bath with temperature $T=H/\pi$. 

This heat bath temperature is twice the Gibbons-Hawking temperature  $T_{\rm GH}$ usually attributed to the cosmological horizon, $T=2T_{\rm GH}$. Moreover,  the electron trajectory is well inside the horizon, since $r_n \ll r_H=1/H$, and thus this process has no relation to the Hawking radiation. However, in Sec. \ref{ConnectionWithHawking} we show that the relation between the temperature of ionization and Gibbons Hawking temperature, $T=2T_{\rm GH}$, is the property of the de Sitter symmetry.

Since the process of ionization takes place well inside the cosmological horizon, one can use the conventional static metric with the gravitational potential $U(r)=-mH^2r^2/2$.\cite{Maxfield2022}
Then the bound state decays by quantum tunnelling from the point $r=0$ to the point $r=r_n$, at which the electron level  $-E_n$ matches the de Sitter gravitational potential  $U(r_n)=-mH^2r_n^2/2$. The radial trajectory $p_r(r)$ is obtained from the classical equation 
\begin{equation}
\frac{{\bf p}^2}{2m}  - \frac{1}{2}mH^2r^2 = -E_n\,,
\label{GravPotential}
\end{equation}
which gives for the radial trajectory:
\begin{eqnarray}
p_r(r)= \sqrt{m^2H^2r^2 -2m E_n}\,.
\label{ElectronTrajectory2}
\end{eqnarray}
The imaginary part of the action at this trajectory again gives Eq.(\ref{IonizationRate}) for the WKB tunneling rate.

\subsection{Decay of composite particles in de Sitter spacetime}
\label{DecayComposite}

The same local temperature $T=H/\pi$ describes the process of  the splitting of the composite particle with mass $m$ into two components with the larger total mass, $m \rightarrow m_1 +m_2>m$, which is also not allowed in the Minkowski vacuum.\cite{Bros2008,Bros2010,Jatkar2012,Volovik2009}

It is instructive to derive the rate of this process using also the semiclassical tunneling picture. For simplicity we consider creation of particles with equal masses, $m_1=m_2>m/2$. The trajectory of each of the two particles with mass $m_1$ moving in the radial direction from the origin at $r=0$ is obtained from equation
\begin{equation}
E(p_r,r) = \sqrt{p_r^2 + m_1^2} + p_rHr= \frac{m}{2}~.
\label{ParticleTraj1}
\end{equation}
We took into account that each of the two particles carries the one half of the energy of the original particle, i.e. $E=m/2$.  The momentum along the trajectory is
\begin{equation}
p_r(r) = \frac{1}{1-H^2r^2} \left[- \frac{m}{2}Hr +\sqrt{ \frac{m^2}{4}-m_1^2+ m_1^2H^2r^2}\right].
\label{ParticleTraj2}
\end{equation}
This momentum is imaginary in the classically forbidden region $r<r_0$, where $r_0=\sqrt{1-m^2/4m_1^2}/H$. This gives the imaginary contribution to the action:
\begin{equation}
{\bf Im}S={\bf Im}\int dr ~p_r(r)=\frac{m_1}{H}\int_0^{r_0}dr \frac{\sqrt{r_0^2-r^2}}{r_H^2-r^2} = \frac{\pi}{4}(2m_1-m).
\label{DecayExponent}
\end{equation}
where as before $r_H=1/H$ is the position of the de Sitter horizon. We must take into account that due to momentum conservation, the two particles tunnel simultaneously in the opposite directions.  Such a coherent process of co-tunneling adds a multiplier of 2 to the argument of the exponent. As a result one obtains the decay rate of the composite particle into two particles, which is again governed by the temperature $T=H/\pi$:
\begin{equation}
\Gamma(1\rightarrow 2) \propto\exp(-4\,{\bf Im}S) =\exp\left(-\frac{\pi(2m_1 -m)}{H}\right)~.
\label{DecayRate2}
\end{equation}

 In general case, when $m_2\neq m_1$, the rate of the decay of the composite particle in the limit $m\gg H$ is determined by the same temperature:
\begin{equation}
\Gamma(1\rightarrow 2) \sim  \exp{\left(-\frac{ \pi(m_1+m_2-m)}{H}\right)}=\exp{\left(-\frac{m_1+m_2-m}{T} \right)}\,.
\label{CompositeRate}
\end{equation}

\subsection{Triplication of particles in de Sitter spacetime}
\label{TtriplicationComposite}

In the same way, massive fermions can be reproduced in the de Sitter environment. For example, the electron at $r=0$  can create an electron-positron pair, with electron and positron moving in opposite directions.  In this process the fermion is triplicated with the rate given by:
\begin{equation}
\Gamma(1\rightarrow 3) \sim  \exp{\left(-\frac{ \pi(3m-m)}{H}\right)}=\exp{\left(-\frac{2m}{T} \right)}\,.
\label{CreationRate}
\end{equation}
Since the created electron and positron move in the opposite directions, they are detected by two different detectors.
Each of the two detectors can detect only single particle, and thus for each detector the radiation rate looks as thermal with the Gibbons-Hawking temperature $T_{\rm GH}=T/2$:
\begin{equation}
\Gamma(1\rightarrow 3) \sim \exp{\left(-\frac{2m}{T} \right)}= \exp{\left(-\frac{m}{T/2} \right)} = \exp{\left(-\frac{m}{T_{\rm GH}} \right)}\,.
\label{CreationRateDouble}
\end{equation}
 However, in this process two particles are created simultaneously and coherently (the so-called co-tunneling process). That is why the temperature, which describes such co-tunneling process of radiation of the pair, is twice  the Gibbons-Hawking temperature. The same connection between the local temperature and the temperature of the Hawking radiation from the cosmological horizon is discussed in Sec. \ref{ConnectionWithHawking}.

It is important that each of three particles (the original electron + electron-positron pair, which is created in the process in Eq.(\ref{CreationRate})) is able to create the other three particles, and then the triplication process continues further. After $n$ replications there will be $3^n$ fermions ($(3^n+1)/2$ electrons and $(3^n-1)/2$ positrons. Such creation of multiple particles from a single particle immersed in the de Sitter environment demonstrates that the de Sitter vacuum is unstable towards the creation of matter. The consequence of such instability will be discussed in Sec. \ref{decay}.

The similar processes take place in the so-called Cosmological Collider,\cite{Maldacena2015,Reece2023} where the new particle created by the Hawking radiation plays the role of the external object which produces the heavy particles.
In this case we have two different physical processes: the Hawking radiation from the cosmological horizon, and the further local process -- the splitting of the created particles, which is determined by the local temperature.

\subsection{Connection between the local and Hawking temperatures}
\label{ConnectionWithHawking}

The local temperature $T=H/\pi$ also determines the process of the Hawking radiation from the cosmological horizon and the Gibbons-Hawking temperature $T_{\rm GH}=T/2=H/2\pi$. The reason is that in the Hawking process, two particles are created coherently (this is the analog of cotunneling): one particle is created inside the horizon, while its partner is simultaneously created outside the horizon.\cite{Parikh2002}  
The rate of the coherent cotunneling of two particles, each with energy $E$, is  $w\propto \exp(-\frac{2E}{T})$. However, the observer who uses the Unruh-DeWitt detector can detect only the particle created inside the horizon. For this observer with limited information, the creation rate $w\propto \exp(-\frac{2E}{T})$ is perceived as 
\begin{equation}
w\propto  \exp\left(-\frac{E}{T/2}\right)=\exp\left(-\frac{E}{T_{\rm GH}}\right)\,,
\label{HawkingRate}
\end{equation}
with the Gibbons-Hawking temperature $T_{\rm GH}=T/2=H/2\pi$.

The similar scenario of doubling the Hawking temperature was suggested for the black hole horizon. \cite{Hooft2022,Hooft2023} In this case the partner of the created particle is in the mirror image of the space-time which replaces the coordinate singularity. Although the existence of such quantum clone of the black hole is problematic, the consideration of the coherent creation of two partners on the two sides of the cosmological horizon is applicable to the de Sitter vacuum.

On the contrary, in the local process of the decay of an atom, which is not related to the cosmological horizon, only single particle (electron) is radiated from the atom. This process is fully determined by the local temperature, 
$w\propto \exp(-\frac{E_n}{T})$. 

\subsection{Two detectors: exited atom vs ionized atom}
\label{ExitedIonized}

Note the main difference between the temperature measured by the observer using the Unruh-DeWitt detector (see e.g. Ref.\cite{Conroy2022} and references therein) and the temperature measured by the observer using ionization of an atom. The ionization process is possible, because the radiated electron moves far away from the atom to position $r=r_0$, where its negative gravitational energy compensates the ionization potential. In the Unruh-DeWitt detector, which corresponds to the two-level atom interacting with a quantum field, the electron in the atom is excited but remains in the same position in the same atom. That is why such excitation of electron may only come either by the Hawking radiation, or by radiation of photons, see Sec. \ref{Photons}.

However, in the de Sitter case there is no difference in temperatures measured by two detectors. If in the Unruh-DeWitt detector experiments, the observer properly interprets the result of the observations according to Eq.(\ref{HawkingRate}), the measured temperature of the Hawking radiation is also $T=2T_{\rm GH}$. So, in spite of different physical principles, both detectors in the de Sitter environment show the same physical temperature, $T=2T_{\rm GH}$.
This also demonstrates the uniqueness of the de Sitter state with its symmetry under the generalized translations.

\subsection{Radiation of photons by atom in de Sitter environment}
\label{Photons}

Instead of the radiation of electron from the atom, one can consider the radiation of photon by the same atom. In both cases the escape of photon or electron from the atom  provides the negative gravitational energy.  That is why these processes, which are prohibited in the Minkowski spacetime, are energetically possible in the de Sitter environment. 
Let us consider the atom in the excited state with energy $E_0 + \epsilon$, where $E_0$ is the ground state energy of the atom. In the de Sitter environment, the radiation of photon is possible even if its energy $cp$ exceeds the energy difference between the excited and the ground state level, i.e. when $cp> \epsilon$.

For the relativistic photon the energy conservation equation (\ref{Classical1}) takes the  form (here we use $c=1$):
\begin{eqnarray}
\sqrt{p_r^2 +p_\perp^2} +p_r Hr = \epsilon\,.
\label{Classical2}
\end{eqnarray}
This gives the following trajectory of photon: 
\begin{eqnarray}
p_r(r) =\frac{1}{1-H^2r^2} 
 \left[-\epsilon Hr \pm \sqrt{\epsilon^2 +p_\perp^2(H^2r^2-1)} \right]\,.
\label{PhotonTrajectory}
\end{eqnarray}
For $cp_\perp > \epsilon$, the integration over the classically forbidden region, $r<r_0$, where $H^2r_0^2 =1 -\epsilon^2/p_\perp^2$, gives the following radiation rate of photon by excited atom:
\begin{equation}
w\sim \exp{ \left(-2\,{\rm Im}\, S\right)}= \exp\left(-\frac{\pi(cp_\perp-\epsilon)}{H}\right) \,\,,\,\, cp_\perp > \epsilon\,.
\label{PhotonRate}
\end{equation}
Since $r_0<1/H$, the trajectory of photon in this process is inside the cosmological horizon, and the radiation rate is again determined by the local temperature $T=H/\pi$.

Eq.(\ref{PhotonRate}) demonstrates, that radiation takes place even if the atom is in the ground state, i.e. when $\epsilon \rightarrow 0$. In this case Eq.(\ref{PhotonRate}) describes the process of radiation of a single photon in the de Sitter environment:
  \begin{equation}
w\sim  \exp\left(-\frac{cp}{T}\right) \,\,,\,\, T=\frac{H}{\pi}\,.
\label{PhotonRate2}
\end{equation}
 In this process, the atom in its ground state serves as an external object, which violates the de Sitter symmetry and thus provides the nonzero matrix element for the radiation of photon.
 This process is certainly different from the Hawking process of radiation.

\subsection{Accelerating detector: is there connection between the local process and Unruh radiation?}
\label{ConnectionWithUnruh}

In the same way as in Sections \ref{ExitedIonized} and \ref{Photons}, one may consider different independent processes related to the accelerating detector. One of them is the Unruh radiation measured by the Unruh-DeWitt detector, which could be linked to the apparent event horizon in the Rindler spacetime and to the corresponding Unruh temperature $T_U=a/2\pi$, where $a$ is acceleration.\cite{Unruh1976,Unruh2019,Unruh2022} The other processes, such as the local process of ionization of the accelerating atom,\cite{Volovik2022e} have no relation to the Rindler horizon.

Let us consider the process of ionization of an atom moving with acceleration. Similar to Eq.(\ref{GravPotential}), the trajectory $p_x(x)$ of the radiated electron is obtained from the classical equation 
\begin{equation}
\frac{{\bf p}^2}{2m}  - max = -\epsilon\,.
\label{UnruhPotential}
\end{equation}
Here, we used the equivalence between the acceleration of the reference frame and the constant gravitational field,  $g=a$; and $\epsilon$ is the ionization potential.

In the WKB approximation and in the limit of slow acceleration, $a\ll \epsilon \ll m$, the classically forbidden trajectory of electron escaping from the atom gives the radiation rate,\cite{Volovik2022e} which is similar to the radiation rate in the electric field:\cite{LandauLifshitzQM}
 \begin{equation}
w\sim \exp\left(-\frac{4\sqrt{2}}{3} \frac{\epsilon}{a} \left(\frac{\epsilon}{m}\right)^{1/2}\right)\,.
\label{IonizationRateUnruh}
\end{equation}
In principle, one can introduce the effective temperature, which is similar to the effective temperature $T_{\rm eff} \sim El$ of hopping electrons in a strong electric field $E$, where $l$ is the localization length.\cite{Shklovskii1992,Shklovskii2024} In our case, the analog of the  localization length is $l=1/\sqrt{m\epsilon}$.  But otherwise the local process of ionization of the accelerated atom is certainly non-thermal. Since $\epsilon\ll m$, the rate of ionization in Eq.(\ref{IonizationRateUnruh}) essentially exceeds the ionization rate in any thermal process, which involves acceleration, $w_{\rm thermal} \sim \exp(-\gamma \epsilon/a)$. This is very different from the ionization of the atom in the de Sitter environment, which looks thermal. 

With the same approach one can also calculate the Unruh radiation of photons using different arrangements of detectors. The state of the art in theoretical understanding of Unruh radiation can be found in Refs. \cite{Gregori2023,Gregori2022,Matsuda2024} and references therein. Our examples demonstrate that the rate of Unruh radiation depends on details of the considered processes and it is not necessarily thermal.
Thermal behavior of the local and global processes in the de Sitter environment is the result of the special de Sitter symmetry.

\section{Thermodynamics of the de Sitter state}
\label{dSthermodynamics}

\subsection{de Sitter symmetry and de Sitter heat bath}
\label{SymmetrySec}

As distinct from the Unruh effect, different arrangements of detectors in the de Sitter environment show the same temperature. This demonstrates the uniqueness of the de Sitter spacetime in producing the thermal bath with local temperature. The reason for that is  that the de Sitter spacetime is homogeneous under the combination of translation and the proper conformal transformations.\cite{Aldrovandi2007,Kamenshchik2008} 
In the PG metric it is the invariance of the de Sitter state under the modified translations, ${\bf r}\rightarrow {\bf r} -e^{Ht}{\bf a}$, which for $H\rightarrow 0$ corresponds to the conventional invariance under translations ${\bf r}\rightarrow {\bf r} -{\bf a}$ in Minkowski spacetime. Due to this combined translational symmetry, all the comoving observers at any point of the de Sitter space observe the same temperature, $T=H/\pi$. That is why one may conclude, that the de Sitter state is the heat bath produced by gravity. 

The uniqueness of the de Sitter thermal state lies in the fact that its temperature does not violate the de Sitter symmetry, and thus does not require the preferred reference frame. This is distinct from thermal state of matter, which always has a preferred reference frame where matter is at rest. As a result, thermal matter violates the de Sitter symmetry, which leads to the heat exchange between the two thermal subsystems, gravity and matter, and finally to the de Sitter decay.

\subsection{From local temperature to local entropy}
\label{FromTtoEntropy}

 The gravity subsystem -- de Sitter quantum vacuum -- has its own temperature and entropy, see also Ref.\cite{Padmanabhan2020}. On the other hand the de Sitter spacetime serves as the thermal bath for matter.  Then the quasi-equilibrium states of the expanding Universe can be described by two different temperatures: the temperature of the gravitational vacuum (temperature of dark energy) and the temperature of the matter degrees of freedom.\cite{Vergeles2023}  
In this section we discuss the pure de Sitter vacuum without the excited matter ignoring for the moment  the thermally activated creation of matter from the vacuum. The excitation and thermalization of matter by the de Sitter thermal bath will be discussed in Sec. \ref{decay}.

If the vacuum thermodynamics is determined by the local activation temperature $T=H/\pi$, then in the Einstein gravity with cosmological constant the vacuum energy density is quadratic in temperature:
\begin{equation}
 \epsilon_{\rm vac}=\frac{3}{8\pi G}H^2=\frac{3\pi}{8G}T^2\,.
\label{dSEnergyDensity}
\end{equation}
This leads to the free energy density of the de Sitter vacuum, $F=\epsilon_{\rm vac} - Ts_{\rm vac}$, which is also quadratic in $T$, and thus the entropy density $s_{\rm vac}$ in the de Sitter vacuum is linear in $T$:
\begin{equation}
s_{\rm vac}= - \frac{\partial F}{\partial T} =\frac{3\pi}{4G}T= 12 \pi^2 KT\,.
\label{dSEntropyDensity}
\end{equation}
Here as before $K=1/16\pi G$ is the gravitational coupling.

The temperature $T$ and the entropy density $s_{\rm vac}$ are the local quantities which can be measured by the local static observer.

\subsection{de Sitter vacuum, Fermi liquid and cosmological constant problem}
\label{FermiLiquidSec}

The equation (\ref{dSEntropyDensity}) demonstrates that the thermal properties of the de Sitter state are similar to that of the non-relativistic Fermi liquid, where the entropy density is also linear in temperature:
\begin{equation}
s_{\rm FL}=\frac{p_F^2}{3v_F}T\,.
\label{FLEntropyDensity}
\end{equation}
The Fermi velocity $v_F$ and Fermi momentum $p_F$ of this cosmological analog of the Fermi liquid are on the order of the speed of light and the inverse Planck length correspondingly, $v_F \sim c$ and 
$p_F\sim M_{\rm Pl}= 1/l_{\rm Pl}$. Of course, this is only a parallel, and one should not identify the de Sitter vacuum with a real Fermi liquid. Although the real Fermi surface may form inside the black hole horizon.\cite{HuhtalaVolovik2002,Zubkov2019}

Since the thermodynamics of the de Sitter state with the thermal energy $\epsilon_{\rm vac}\propto T^2$ is similar to the thermodynamics of the Fermi liquid, let us try to exploit this connection. One of the directions is the Sommerfeld law in Fermi liquid, which states that the entropy per one atom of the Fermi liquid is 
\begin{equation}
S=\frac{s_{\rm FL}}{n_{\rm FL}}\sim \frac{T}{E_F}\,,
\label{FLEntropyDensity}
\end{equation}
 where $n_{\rm FL} \sim p_F^3$ is the density of atoms in the Fermi liquid and $E_F$ is Fermi energy. 
 
 We do not know what are  the "atoms of the vacuum", but from Eq.(\ref{dSEntropyDensity}) it follows that the entropy density of the vacuum  
$s_{\rm vac} \sim T/l_{\rm Pl}^2 \sim (T/M_{\rm Pl})/l_{\rm Pl}^3$, where $l_{\rm Pl}$ is the Planck length and $M_{\rm Pl}$ is the Planck energy.
This suggests that the density of the "atoms of the vacuum" is $n_{\rm Pl} \sim 1/l_{\rm Pl}^3$ and entropy per  "atom of the vacuum" is:
\begin{equation}
S= \frac{s_{\rm vac}}{n_{\rm Pl}} \sim s_{\rm vac} l_{\rm Pl}^3 \sim \frac{T}{M_{\rm Pl}}  \,.
\label{Sommerfeld}
\end{equation}
  Eq.(\ref{Sommerfeld}) is the full analog of the Sommerfeld law for Fermi liquid. This analogy also suggests that the corresponding density of states in the quantum vacuum (the analog of density of states at the Fermi level $N_F\sim mp_F$  in Fermi liquids) is $N_{\rm Pl} \sim M_{\rm Pl}^2$. For bosonic and fermionic degrees of freedom of this quantum vacuum the density of states is $N_{\rm Pl} =9/4\pi G$ and $N_{\rm Pl} =9/2\pi G$ correspondingly, which can be compared with the value  $N_{\rm Pl} =3\pi/G$ suggested in Ref. \cite{Chu2023}. This huge density of states leads to a very large entropy of the de Sitter state even at very small temperature of the vacuum.
 
 So,  the quantum vacuum looks as some specific form of the relativistic Fermi liquid. However, some consequences are the same as in the non-relativistic Fermi liquid. In particular,  in the full equilibrium at $T=0$ and in the absence of external pressure, $P=0$, the energy density of the non-relativistic Fermi liquid  is exactly zero, $\epsilon(n) -\mu n=Ts -P=0$. This follows solely from thermodynamics, and is valid for any macroscopic system in its ground state. In all such systems, the contribution of the zero-point energies of collective modes (phonons, fermionic quasiparticles, magnions, etc.) is exactly compensated by the microscopic (atomic) degrees of freedom. This thermodynamic property does not depend on the microscopic physics, and is easily checked in condensed matter systems, where we know both the micro and macro physics. That is why it is not surprizing, that for the relativistic quantum vacuum, the vacuum energy density is zero if the following conditions are fulfiled: there is the full thermodynamic equilibrium; the temperature $T=0$;  there is no matter (no quasiparticles);  and there is no external pressure, $\epsilon_{\rm vac} (H=0, T=0)=-P_{\rm vac}=0$. This demonstrates, that thermodynamics, which is insensitive to the microscopic structure of the vacuum (ground state), solves the main cosmological constant problem.
 The other cosmological constant problems are discussed in Sec. \ref{CCProblemsSec}.

\subsection{Hubble volume entropy vs entropy of the cosmological horizon}
\label{HubbleVolume}

Using the entropy density in Eq.(\ref{dSEntropyDensity}), one may find the total entropy of the Hubble volume $V_H$ -- the volume  surrounded by the cosmological horizon with radius $R=1/H$:
\begin{equation}
S_{\rm bulk}=s_{\rm vac}V_H=\frac{4\pi R^3}{3} s _{\rm vac}= \frac{\pi}{GH^2}= \frac{A}{4G} =S_{\rm hor}\,,
\label{dSEntropy}
\end{equation}
where $A$ is the horizon area. 

The Hubble-volume entropy coincides with the entropy attributed to the cosmological horizon as suggested by Gibbons and Hawking. However, Eq.(\ref{dSEntropy}) demonstrates that the thermodynamic entropy comes from the local entropy of the de Sitter quantum vacuum, rather than from the hypothetical horizon degrees of freedom. In this hypothesis it assumed that there are the horizon microstates, which are concentrated in the region of the Planck length $l_{\rm Pl}=1/M_{\rm Pl}$ in the vicinity of horizon, and this leads to the area law for the total entropy inside the horizon, see e.g. Ref. \cite{Padmanabhan2010}. The local thermodynamics in the de Sitter state  leads to the same area law, but without assumptions on the spatial distribution of the degrees of freedom. 

The connection between the bulk entropy of the region inside the horizon due to the local vacuum thermodynamics and the surface entropy of the horizon may have the  holographic origin, when the horizon is considered as null surface.\cite{Padmanabhan2010} This may be the reason why the presence of the horizon allows us to know the total entropy of the Hubble volume without knowledge on the distribution of the local entropy inside the horizon.

However, it appears that the bulk-surface correspondence in Eq.(\ref{dSEntropy}) is valid only in the ($3+1$)-dimensional spacetime. In the general $d+1$ dimension of spacetime, the same approach gives the factor $(d-1)/2$ in the relation between the entropy of the Hubble volume and the Gibbons-Hawking entropy of the cosmological horizon, $S_{\rm bulk} =\frac{d-1}{2} \,S_{\rm hor}$. This may add to the peculiarities of the $d=3$ space dimension,\cite{Barrow1983} where in particular the mass dimension of the gravitational coupling, $[K]=d-1$, coincides with the mass dimension of curvature, $[{\cal R}]=2$.
The same concerns such pair of thermodynamically conjugate variables as electric field with the mass dimension $[E]=2$ and the electric induction with mass dimension 
$[D]=d-1$. Their dimensions also coincide only for $d=3$. Discussion of the natural dimensions of physical quantities in $d=3$ can be found in Ref. \cite{Volovik2022f}.

\section{Thermodynamics from the heat transfer in the multi-metric gravity ensemble}
\label{HeatTransferSec}

 \subsection{Multi-metric gravity}
\label{MultiMetricSec}

As we know, the main source of the emergent thermodynamics is the heat exchange between the bodies or between the systems, see also Ref.\cite{Strasberg2024}. Following this rule, we can consider the de Sitter thermodynamics from the point of view of the heat transfer between different cosmological objects or different Universes. 

The heat exchange can be discussed in the frame of the so-called multi-metric gravity, see Ref.\cite{Wood2024} and references therein. 
The corresponding model action of the whole system can be written as the sum of actions of the sub-systems in the same coordinate spacetime:
\begin{equation}
S=-\int d^4 x \sum_{n=1}^N {\cal L}_n \,\,,\,\, {\cal L}_n=\sqrt{-g_{(n)}} \left(K_n {\cal R}\{g_{\mu\nu (n)}\}+\Lambda_n\right)\,.
\label{RandomAction}
\end{equation}
Then the Universe can be seen as the system of $N$ sub-Universes, each with its own gravitational coupling $K_n$, cosmological constant $\Lambda_n$ and metric $g_{\mu\nu (n)}$.

Following Froggatt and Nielsen\cite{FroggattNielsen1991} one can introduce $N$ independent tetrad fields $e^{a(n)}_\mu$ for $N$ fermionic species. In this multi-tetrad gravity one has the ensemble of the gravitational actions:
\begin{eqnarray}
{\cal L}_n=e_{abcd}\left(K_n\,R^{ab(n)}\wedge e^{c(n)}\wedge e^{d(n)}+
\Lambda_n\,e^{a(n)}\wedge e^{b(n)} \wedge e^{c(n)}\wedge e^{d(n)}\right) \,,
\label{TetradAction}
\end{eqnarray}
and the corresponding ensemble of actions for the fermionic species:
\begin{equation}
S_M= e_{abcd}\int  \sum_n \Theta^{a(n)}\wedge e^{b(n)} \wedge e^{c(n)}\wedge e^{d(n)}
\,,
\label{Action1}
\end{equation} 
\begin{equation}
 \Theta^{a(n)} = \frac{i}{2} \left[ \bar\Psi^{(n)} \gamma^a D_\mu \Psi^{(n)} - D_\mu \bar\Psi^{(n)}\gamma^a \Psi^{(n)}\right] dx^\mu
\,.
\label{Action2}
\end{equation}
This can be extended to the multi-f\"unfbein gravity, where instead of the tetrad fields the Dirac fermions are described by the rectangular vielbein (f\"unfbein).\cite{ObukhovVolovik2024}

Gravity with multiple tetrad fields may also come from the Akama-Diakonov-Wetterich theory, \cite{Akama1978,Wetterich2004,Diakonov2011,VladimirovDiakonov2012,VladimirovDiakonov2014,Sindoni2012,ObukhovHehl2012,Volovik1990,WeiLu2024} where the tetrads are formed as composite objects -- the bilinear combinations of the fundamental fermionic fields:
\begin{equation}
e^{a(n)}_\mu=\left< \Theta^{a(n)}  \right>
\,.
\label{bilinear}
\end{equation}
In this approach, the metric is the quartet of fermions. In principle, the so called vestigial gravity can be realized, in which the bilinear combination of fermions in Eq.(\ref{bilinear}) is zero,  $e^{a(n)}_\mu=0$, while the metric -- the  quartet of fermions -- is nonzero:\cite{Volovik2024q}
\begin{equation}
g_{\mu\nu(n)}=\eta_{ab}\left< \Theta_\mu^{a(n)}  \Theta_\nu^{b(n)} \right>
\,.
\label{quartet}
\end{equation}
On the levels of particles, the vestigial gravity acts with different strength on fermions and bosons.
In principle, it is not excluded that such gravity can be formed at an early stage of the development of the Universe.

\subsection{Heat exchange in multi-metric gravity}
\label{TwoHeatTransferSec}

The heat exchange between the sub-Universes leads to their equilibration with formation of the common expansion rate and thus the common temperature. 
We consider first the system of two sub-Universes, assuming that in each of them the entropy of horizon obeys the area law, and show that the maximum entropy corresponds to the situation in which both states acquire  the same expansion rate. 

In the de Sitter state, which is determined by the cosmological constant, the equation of state for the vacuum energy  is  $\epsilon_{\rm vac}=-P_{\rm vac}$. The total vacuum energy  is proportional to  the volume $V$ of the system, if we assume that the volume $V$ is much larger than the Hubble volume, $V\gg V_H$, so that the boundary terms are not important.
Then we have:
\begin{equation}
E_V= \epsilon_{\rm vac} V=6KH^2V\,.
\label{TotalEnergy}
\end{equation}

Let us assume that the bulk-surface correspondence is valid, i.e. the entropy of the Hubble volume $V_H$ is equal to the Gibbons-Hawking entropy of cosmological horizon, $S_{V_H}=S_{\rm hor}=4\pi KA$.
Then the total entropy $S_V$ in the volume $V\gg V_H$ can be obtained from the entropy of the Hubble volume $V_H$:
\begin{equation}
S_{V}=S_{\rm hor} \frac{V}{V_H} =12 \pi KHV\,.
\label{TotalEntropy}
\end{equation}

Let us now consider two de Sitter sub-states  with different values of the gravitational coupling, $K_1$ and $K_2$, and different values of the Hubble parameter, $H_1$ and $H_2$:
\begin{eqnarray}
S=-\int d^4 x\sqrt{-g_{(1)}}  \left(K_1 {\cal R}\{g_{\mu\nu (1)}\}+\Lambda_1\right) -\int d^4 x\sqrt{-g_{(2)}}  \left(K_2 {\cal R}\{g_{\mu\nu (2)}\}+\Lambda_2\right)\,.
\label{TwoUniverses}
\end{eqnarray}
 This corresponds to the higher dimensional analog of the bilayer graphene,\cite{Parhizkar2022} where two 2+1 dimensional Universes are in the neighbouring layers of the 3+1 spacetime. In this interpretation we have two 3+1 dimensional Universes in the neighbouring layers in the 4+1 space.  
 
 The total energy and total entropy of two layers are (if the interaction between the layers is neglected):
\begin{equation}
E_V=E_1+E_2=6(K_1H_1^2 + K_2H_2^2)V  \,,
\label{TwoUniversesE}
\end{equation}
\begin{equation}
S_V=S_1+S_2=12 \pi (K_1H_1 + K_2H_2)V \,.
\label{TwoUniversesS}
\end{equation}

Let us now allow for the energy exchange (the heat exchange) between these two sub-Universes (analogs of the two layers of graphene). This exchange can be realized by the matter field, which interacts with both metrics. It leads to the variations of the Hubble parameters $H_1$ and $H_2$ at fixed $E_V$. If we ignore the thermalization  of matter by de Sitter environment, the heat exchange will finally produce the equilibrium state with the maximum entropy $S$, in which the Hubble parameters become equal:
\begin{equation}
H_1^2=H_2^2=\frac{E_V}{6(K_1+K_2)V}\equiv H^2\,.
\label{HubbleEquilibrium}
\end{equation}
The equilibration of the Hubble parameters demonstrates that the Hubble parameter (with some numerical factor)  plays the role of the temperature of the de Sitter Universe.

 The temperature of the de Sitter Universe can be obtained by variation over the Hubble parameter:
 \begin{eqnarray}
\frac{1}{T_1}=\frac{dS_1}{dE_1}=\frac{dS_1/dH_1}{dE_1/dH_1}=\frac{\pi}{H_1} \,,
\nonumber
\\
 \frac{1}{T_2}=\frac{dS_2}{dE_2}=\frac{dS_2/dH_2}{dE_2/dH_2}=\frac{\pi}{H_2}\,,
\label{Temperature}
\end{eqnarray}
with $T_1=T_2=H/\pi$ in equilibrium.

In case of the arbitrary number $N$ of the sub-Universes, the heat exchange between them leads to the equilibrium state of the Universe in which all the sub-Universes coherently expand with the same rate $H$, i.e., with the same de Sitter metric in all the subsystems. In this equilibrium Universe, the gravitational coupling $K$ is equal to the sum of the individual couplings in the sub-Universes and the vacuum energy density is equal to the sum of energy densities of subsystems:
 \begin{equation}
E_V=6KH^2V\,\,,\,\, K=\sum_n K_n \,\,,\,\, \Lambda=\sum_n \Lambda_n\,.
\label{MultiHubble}
\end{equation}
All the substates acquire the same temperature in equilibrium, $T_n=T=H/\pi$.

\subsection{Thermodynamics from the multi-metric ensemble}
\label{MultiEnsembleSec}

Let us remind that in the above approach we used the bulk-horizon correspondence $S_{V_H}=S_{\rm hor}=4\pi KA$, which finally leads to the equilibrium Universe with temperature $T=H/\pi$.
Let us now consider the thermodynamics of the whole de Sitter system without assumption about the entropy of the cosmological horizon. For that we consider the statistical ensemble of $N$ de Sitter sub-Universes with random Hubble parameters $H_n$. This is the extension of the multi-metric gravity to the statistical ensemble with the randomly distributed parameters $K_n$ and $\Lambda_n$. 

For large $N$, the random distribution of the parameters results in the exponential behaviour of the distribution functions, $w_n\propto \exp(-E_n/T)$, with the same parameter $T$ for all subsystems. 
As in the statistical ensemble of atoms, where the temperature of the system is determined by the physical processes, the temperature of the ensemble of the sub-Universes is also determined by the physical processes. In our case it is the behaviour of matter (atom) in the de Sitter environment, which gives $T=H/\pi$. This connection between $T$ and $H$  is rather natural.
Both, the parameter $T$, which plays the role of temperature, and the Hubble parameter $H$ are the quantities which in equilibrium become common for all the subsystems in the ensemble, and they have the same dimension of inverse time, $[T]=[H]= [1/t]$. 

The physical temperature in turn gives rise to the total entropy and to the local entropy, $S_V=\sum_n S_n=12 \pi KHV=s_{\rm loc}V$. So, in this scenario the
de Sitter  entropy comes from a set of many randomly distributed subsystems with the expansion rates $H_n$. Due to the heat exchange at fixed total energy $E_V$ these states are organized in the equilibrium thermal state, which corresponds to the coherent de Sitter expansion of the whole system. The coherence due to thermalization may explain the horizon problem, i.e. why the causally-disconnected regions of the CMB are in thermal equilibrium.

\subsection{Regularization vs thermalization}
\label{Regularization}

The multi-metric ensemble may include the ensemble of $N$ species of Weyl or Dirac fermions. At large $N$, all tetrads in the random ensemble approach the same value, $e^{a(n)}_\mu \rightarrow e^a_\mu$, and thus in the equilibrium state all fermionic species experience the same geometry. In Refs. \cite{FroggattNielsen1991,ChadhaNielsen1983} the formation of the common Lorentz invariance for different fermionic species was also considered. But this was achieved by the renormalization group effect in the infrared limit, instead of thermalization. This suggests the possible connection between  renormalization and thermalization.

One may expect that the heat exchange between subsystems leads not only to the coherence of the de Sitter states, but to the general coherence of the metric fields, when the metric fields $g_{\mu\nu(n)}$ of the subsystems become equal, thus forming the common metric $g_{\mu\nu}$.
If this is true, this could be a kind of the thermodynamic gravity, but without using the holographic principle.

\subsection{Coherence vs thermalization}
\label{Coherence}

At first glance this formation of the coherent de Sitter expansion from the ensemble of the random microstates looks similar to the formation of the Bose-Einstein condensate (BEC) of magnons.\cite{BunkovVolovik2013} The magnon BEC represents the coherent precession of all spins, which results from the incoherent precessions of the individual spins with random frequencies $\omega_n$ in the local magnetic fields. The originally random frequencies correspond to the random Hubble parameters $H_n$ in different sub-Universes. The coherence of precession develops due to the spin currents between the regions of the local precessions -- the analog of the heat exchange between the sub-Universes. The formed common frequency $\omega$ of the coherent precession corresponds to the formed common Hubble parameter $H$ of the large Universe. 
Moreover, they have the same dimension of the inverse time, $[H]=[\omega]=[1/t]$, while the dimensionless  number of magnons (or, which is the same, the dimensionless spin projection $S_z$) corresponds to the total entropy of de Sitter which is also dimensionless. 

However, the source of the coherence is the exchange of spins instead of the exchange of the energies. As a result the formation of the coherent state of spin precession is due to minimization of the total energy $E$ at fixed projection $S_z$ of the total spin on magnetic field. This leads to the common frequency of precession. In the magnon BEC interpretation,  the common frequency plays the role of the chemical potential $\mu$ for magnons, which becomes constant in space due to the exchange of magnons between the regions. That is why this process is quite opposite to the formation of the common temperature, where the total energy is fixed, while the total entropy reaches its maximum value. 

Anyway, in all the cases the parameters, which are the same in all subsystems in equilibrium -- common frequency $\omega$, common chemical potential $\mu$, common temperature $T$, common angular velocity $\boldsymbol\Omega$ and now also the common Hubble parameter $H$ -- all of them have the same dimension of inverse time, $[T]=[H]=[\mu]=[\omega]=[\Omega]=[1/t]$. 
In Sec. \ref{AnomalySec} it will be shown that the Hubble parameter $H$ enters the chiral anomaly effects together with the other thermodynamic variables: $T$, $\mu$ and 
$\boldsymbol\Omega$. All this supports the thermodynamic nature of the Hubble parameter.

\subsection{de Sitter contribution to chiral anomaly}
\label{AnomalySec}

As the thermodynamic quantity, the Hubble parameter $H$ participates in different thermodynamical effects. We consider this using as an example the chiral vortical effect -- the appearance of the chiral current in fermionic systems in the presence of rotation. For Dirac fermions in the flat spacetime, the chiral current ${\bf j}_A $ contains the following contributions from temperature, chemical potential and angular velocity:\cite{Vilenkin1979,Vilenkin1980,Kharzeev2016,Stone2018,Abramchuk2024} 
\begin{equation}
 {\bf j}_A = \left( \frac{T^2}{6} + \frac{\mu^2}{2\pi^2}  + \frac{\Omega^2}{24\pi^2}\right)\boldsymbol\Omega\,.
\label{ChiralCurrentT}
\end{equation}

In the de Sitter state the gravitational thermodynamical variable -- the curvature ${\cal R}$ -- is added.\cite{Flachi2018,Khakimov2024} According to Ref.  \cite{Khakimov2024} the gradient expansion gives the following contribution to the chiral current, which is very similar to the contributions of other thermodynamic quantities in Eq.(\ref{ChiralCurrentT}):
\begin{equation}
 {\bf j}_A =  \frac{H^2}{8\pi^2} \boldsymbol\Omega
\,.
\label{ChiralCurrentR}
\end{equation}
The same term is obtained for the anti-de Sitter spacetime where $H^2<0$.\cite{Ambrus2021}
On the other hand, in Ref.\cite{Flachi2018} the curvature term comes from the shift of the fermionic mass gap, $-M^2 \rightarrow -(M^2+H^2)$, and this leads to the Eq.(\ref{ChiralCurrentR}) with opposite sign.

Comparison of Eq.(\ref{ChiralCurrentR}) with Eq.(\ref{ChiralCurrentT}) demonstrates that there is the difference between the contribution $T^2/6$ in Eq.(\ref{ChiralCurrentT}), which comes from the temperature of matter, and the contribution  $T^2/8$ from the temperature $T=H/\pi$ of the de Sitter environment in Eq.(\ref{ChiralCurrentR}). However, it is possible that the more rigorous calculations (beyond the gradient expansion, with the proper conservation laws and with the proper limit cases) can modify the coefficient in Eq.(\ref{ChiralCurrentR}). It is not excluded that these two contributions may cancel each other when the matter and the gravitational background are in equilibrium and thus have the same temperature. Such cancellation of the currents generated by two different mixed gravitational anomalies in rotating chiral liquid under the full equilibrium was found in Ref. \cite{NissinenVolovik2022}. This supports the Bloch theorem on the absence of the total current in equilibrium, see also Refs. \cite{Yamamoto2015,Volovik2017,Zubkov2018}.

Anyway, the participation of the curvature ${\cal R}=-12H^2$ in the thermodynamics of the chiral anomaly demonstrates the uniqueness of the de Sitter state in providing the contributions of gravitational variables to different thermodynamic effects  together with the traditional thermodynamic variables, such as temperature, chemical potential, angular velocity, electric and magnetic fields.

\section{Thermodynamics of de Sitter state and $f(R)$ gravity}
\label{dSthermodynamicsSec}

\subsection{Thermodynamic variables in $f(R)$ gravity}
\label{VariablesSec}

Here we consider the thermodynamics of the de Sitter state in terms of the gravitational variables and the corresponding modification of the thermodynamic Gibbs-Duhem relation for the quantum vacuum. 

The conventional vacuum pressure $P_{\rm vac}$ obeys the equation of state $w=-1$ and  enters the energy momentum tensor of the vacuum medium in the form:
\begin{equation}
T^{\mu\nu}= \Lambda g^{\mu\nu} = {\rm diag}( \epsilon_{\rm vac},  P_{\rm vac},   P_{\rm vac}, P_{\rm vac}) \,\,, \,\,
P_{\rm vac}=-\epsilon_{\rm vac}\,.
\label{EnergyMomentum}
\end{equation}
In the de Sitter state the vacuum pressure  is negative, $P_{\rm vac}=-\epsilon_{\rm vac}<0$. 

Due to the linear dependence of the de Sitter entropy density on temperature this pressure $P_{\rm vac}$ does not satisfy the standard thermodynamic Gibbs-Duhem relation, $Ts_{\rm vac}=  \epsilon_{\rm vac}+ P_{\rm vac}$, because the right hand side of this equation is zero.
The reason for that is that in this equation we did not take into account the gravitational degrees of freedom of quantum vacuum. Earlier, it was shown that gravity contributes to thermodynamics with a pair of thermodynamically conjugate variables:  the gravitational coupling $K=\frac{1}{16\pi G}$ and the scalar Riemann curvature ${\cal R}$, see Refs.\cite{KlinkhamerVolovik2008c,Volovik2022,Volovik2020}. The  contribution of the term  $K{\cal R}$  to thermodynamics is similar to the work density.\cite{Hayward1998,Hayward1999,Jacobson1995,Odintsov2023a} 

The quantities $K$ and ${\cal R}$  can be considered as the local thermodynamic variables, which in
condensed matter physics are similar to temperature, pressure, chemical potential, number density, spin density, etc. Indeed, since the de Sitter spacetime is maximally symmetric, its local structure is characterized by the scalar curvature alone, while all the other components of the Riemann curvature tensor are expressed via ${\cal R}$:
 \begin{equation}
R_{\mu\nu\alpha\beta} = \frac{1}{12}\left(g_{\mu\alpha} g_{\nu\beta} - g_{\mu\beta} g_{\nu\alpha}\right){\cal R}\,.
\label{CurvatureTensor}
\end{equation}
That is why the scalar Riemann curvature as the covariant quantity naturally serves as one of the
thermodynamical characteristics of the macroscopic matter.\cite{Pronin1987,Pronin1995}

Another argument is related to the so-called Larkin-Pikin effect.\cite{LarkinPikin1969}
 This is the jump in the number of degrees of freedom, when the fully homogeneous state is considered. One has the extra parameters,  which are space independent, but participate in thermodynamics. \cite{Polyakov1991,KlinkhamerVolovik2008,Polyakov2022}
The same concerns the constant electric
and magnetic fields {\it in vacuo}, which together add three more degrees of freedom. These constant fields are mutually independent,
in contrast to the spacetime-dependent fields connected by the Maxwell equations.\cite{KlinkhamerVolovik2008}
  The scalar curvature ${\cal R}$ in the de Sitter vacuum, which is constant in space-time, also serves as such thermodynamic parameter. Then the gravitational coupling $K=df/d{\cal R}$ serves as the analog of the chemical potential, which is constant in the full equilibrium.

\subsection{Gibbs-Duhem relation in $f(R)$ gravity}
\label{GibbsExtensionSec}

The new thermodynamic variables, $K$ and ${\cal R}$, which come from the gravity, and Eq. (\ref{dSEntropyDensity}) for the entropy density allow us to restore the Gibbs-Duhem relation for de Sitter vacuum in the following form:
 \begin{equation}
Ts_{\rm vac}=  \epsilon_{\rm vac}+ P_{\rm vac} -K{\cal R}\,.
\label{GibbsDuhem}
\end{equation}
This equation is obeyed, as can be checked on example of the Einstein gravity, where $Ts_{\rm vac}= 12 \pi^2 KT^2=12KH^2$,  using the equations
$\epsilon_{\rm vac}+ P_{\rm vac}=0$ and ${\cal R}=-12H^2$. This supports the earlier proposal that $K$ and ${\cal R}$ can be considered as the thermodynamically conjugate variables.\cite{Volovik2022,Volovik2020}

The Eq.(\ref{GibbsDuhem}) can be also written using the effective vacuum pressure, which absorbs the gravitational degrees of freedom:
\begin{equation}
P= P_{\rm vac} -K{\cal R} \,.
\label{EffectiveP}
\end{equation}
Then the conventional  Gibbs-Duhem relation is restored:
\begin{equation}
Ts_{\rm vac}=  \epsilon_{\rm vac}+ P\,.
\label{EffectiveGibbs}
\end{equation}

The equation (\ref{EffectiveGibbs}) is just another form of writing the Gibbs-Duhem relation (\ref{GibbsDuhem}). But it allows to make different interpretation of the de Sitter vacuum state. The introduced effective de Sitter pressure $P$ is positive, $P=\epsilon_{\rm vac}>0$, and satisfies equation  of state $w=1$, which is similar to matter with the same equation of state. As a result, due to the gravitational degrees of freedom, the de Sitter state has many common properties with the non-relativistic Fermi liquid, where the thermal energy is proportional to $T^2$, and also with the relativistic stiff matter with $w=1$ introduced by Zel'dovich.\cite{Zeldovich1962}

\subsection{Entropy of cosmological horizon in terms of effective gravitational coupling}
\label{EntropyKSec}

Let us now show that the holographic bulk-surface correspondence remains valid also in the $f( {\cal R})$ gravity, i.e. that the entropy of the Hubble volume coincides with the entropy attributed to the cosmological horizon, $S_{\rm hor}=4\pi KA$.

In the $f( {\cal R})$ gravity the action is:
\begin{equation}
S=-\int d^4 x \sqrt{-g} f( {\cal R})\,.
\label{action}
\end{equation}
In the equilibrium de Sitter state the curvature is determined by the Einstein equations obtained by variation of the action (\ref{action}):
\begin{equation}
2 f( {\cal R})={\cal R}\frac{df}{d {\cal R}}\,.
\label{dS}
\end{equation}
The corresponding vacuum energy density $\epsilon_{\rm vac}=- P_{\rm vac}$ in the $f( {\cal R})$ gravity is:
 \begin{eqnarray}
 \epsilon_{\rm vac}=f({\cal R}) -K{\cal R} \,\,, \,\, K=\frac{df}{d {\cal R}}   \,.
\label{GibbsDuhemGeneral2}
\end{eqnarray}
This shows that the gravitational coupling $K$ is the natural definition of the variable, which is thermodynamically conjugate to the curvature ${\cal R}$, while $\epsilon_{\rm vac}$ serves as the corresponding thermodynamic potential.
The Gibbs-Duhem relation for the de Sitter states  in the $f( {\cal R})$ gravity has the conventional thermodynamic form:
 \begin{eqnarray}
 Ts_{\rm vac}=  \epsilon_{\rm vac}+ P_{\rm vac} -K{\cal R} \,.
\label{GibbsDuhemGeneral1}
\end{eqnarray}

Let us now use the fact, that the local temperature $T=H/\pi$ of the equilibrium de Sitter state is the geometric property of the de Sitter, being  fully determined by the PG metric in Eq.(\ref{PG1}). This temperature, which in particular regulates the process of the ionization of an atom in the de Sitter environment, does not depend on the function $f( {\cal R})$. Using this local temperature one obtains from Eq.(\ref{GibbsDuhemGeneral1}) the local entropy,
$s_{\rm vac}=-K{\cal R}/T$, where ${\cal R}=-12H^2=-12\pi^2T^2$.
Then the total entropy of  the Hubble volume $V_H$ is given by the same Eq.(\ref{dSEntropy}) as in the Einstein gravity: 
 \begin{equation}
S_{\rm bulk}=s_{\rm vac}V_H=4\pi KA=S_{\rm hor} \,.
\label{dSEntropyGen}
\end{equation}
But now $K$ is the effective gravitational coupling in Eq.(\ref{GibbsDuhemGeneral2}). 

This generalization of the Gibbons-Hawking entropy was discussed in Refs.\cite{Odintsov2005,KlinkhamerVolovik2008c,Brustein2009,Chao-QiangGeng2019}. But here it was obtained using the local thermodynamics of the de Sitter vacuum.
This demonstrates that the local thermodynamics of the de Sitter vacuum is valid also for the 
$f({\cal R})$ gravity.  The effective gravitational coupling $K$ serves as one of the thermodynamic variable of the  local thermodynamics. This quantity plays the role of the chemical potential, which is thermodynamically conjugate to the curvature ${\cal R}$, and it is constant in the thermodynamic equilibrium state of de Sitter spacetime.

\subsection{Example of quadratic gravity}
\label{QuadraticSec}

For illustration,  we consider the simple example of the $f({\cal R})$ gravity and the corresponding modification of the gravitational coupling $K$ in the de Sitter environment.
In the conventional Einstein gravity, where $f({\cal R})=K_0{\cal R} + \Lambda$, the de Sitter state has the equilibrium value of the curvature, ${\cal R}_0=-2\Lambda/K_0= -12 H^2$.  Let us add the quadratic term to the Einstein action:\cite{Odintsov2005,KlinkhamerVolovik2008c}
\begin{equation}
f({\cal R})=K_0{\cal R} -\frac{1}{2}\epsilon_G {\cal R}^2+ \Lambda\,.
\label{Rsquare1}
\end{equation}
Here $\epsilon_G$ is the dimensionless parameter. In electrodynamics, this parameter corresponds to such parameters as dielectric constant, magnetic permeability and inverse fine structure constant.  These parameters contain the logarithm, $\ln \frac{M_{\rm Pl}^4}{{\bf B}^2}$. This is the running coupling, in which the ultraviolet cut-off is given by Planck mass, while the infrared cut-off is provided by the magnetic field  ${\bf B}$. The coefficient of the logarithmic term depends on the number of massless fermionic and bosonic species. In gravity, there is also the logarithmic correction $\ln \frac{M_{\rm Pl}^4}{{\cal R}^2}$ to $\epsilon_G$, were the infrared cut-off is provided by the Hubble parameter. But here we ignore the logarithmic contributions for simplicity.

In the analogy with electrodynamics, the parameter $K_0$ corresponds to the spontaneous magnetization or to the spontaneous polarization, which break the corresponding discrete $T$ and $P$ symmetries. In gravity,  the corresponding broken discrete symmetry is the symmetry with respect to transformation ${\cal R}\rightarrow -{\cal R}$. The possible origin of such discrete symmetry\cite{Volovik2024Z4} is the symmetry under the coordinate transformation $x^\mu \rightarrow ix^\mu$ (the complex metric was also considered in Ref.\cite{Bondarenko2022}). Under this operation the de Sitter state is transformed to the anti-de Sitter, see also Ref. \cite{Bzowski2024}. This is different from the time reversal symmetry operation, which connects the black and white holes in Sec. \ref{WHentropy}, 
and the expanding and contracting de Sitter states in Sec. \ref{ContractingDSSec}. 

The equilibrium curvature ${\cal R}_0$ in the $f({\cal R})$ gravity in Eq.(\ref{Rsquare1}) can be obtained from Eq.(\ref{dS}):
\begin{eqnarray}
{\cal R}_0=-2\frac{\Lambda}{K_0} = -12 H^2\,,
\label{Rsquare2}
\end{eqnarray} 
It is the same as in Einstein gravity, because the quadratic terms in Eq.(\ref{dS}) are cancelled.
 The equilibrium value of the effective gravitational coupling $K$ is: 
\begin{eqnarray}
 K=\frac{df}{d {\cal R}}\big\vert_{{\cal R}={\cal R}_0}=K_0+ 2 \epsilon_G\frac{\Lambda}{K_0}\,.
\label{Rsquare4}
\end{eqnarray}
This modified gravitational coupling $K$ determines the local entropy  $s_{\rm vac}$, which follows from Eq.(\ref{GibbsDuhemGeneral1}). As a result, the entropy of the Hubble volume in Eq.(\ref{Rsquare3}), which we identify with the entropy attributed to the horizon  $S_{\rm hor}$, is also determined by the modified coupling $K$:    
   \begin{eqnarray}
S_{\rm hor}=s_{\rm vac}V_H=4\pi KA  \,.
\label{Rsquare3}
\end{eqnarray}

The local and global entropies  change sign for $K<0$, while the cosmological expansion is still described by the de Sitter metric. However, the negative $K$ requires the negative parameter $\epsilon_G<0$, which marks the instability of such de Sitter vacuum.\cite{Odintsov2005}

\section{From de Sitter thermodynamics to de Sitter decay}
\label{decay}

\subsection{de Sitter state as thermal bath for matter}
\label{ThermalBathSec}

The validity of the holographic connection between the bulk and surface entropies in the extension of the thermodynamics to the $f({\cal R})$ gravity also supports the idea that the de Sitter vacuum is the thermal state with the local temperature $T=H/\pi$. Such gravitational temperature, which is twice the Hawking temperature, has been also obtained in the particular de Sitter limit, when the relativistic and non-relativistic matter tend to zero (see the footnote 2 on page 4 in Ref.\cite{Boyle2024}). This demonstrates that in the conventional approaches to the de Sitter thermodynamics, where the Euclidean
time is used, the results may depend on the choice of the order of limits.

The nonzero local temperature of the gravitational vacuum shows that the de Sitter vacuum is locally unstable towards the creation of matter, if some matter (such as atom or electron) is originally present, see Section \ref{TtriplicationComposite}.
This is distinct from the mechanism of creation of the pairs of particles by Hawking radiation from the cosmological horizon, which may or may not lead to the decay of the vacuum energy. There are still controversies concerning the stability of the de Sitter vacuum caused by the Hawking radiation, see e.g., Refs.
\cite{Kamenshchik2022,Polyakov2012,Polyakov2022}  and references therein.

\subsection{de Sitter decay due to thermalization of matter by de Sitter heat bath}
\label{DeacyByThermalizationSec}

To describe the decay of the vacuum due to creation and the further thermalization of matter, the extension of the Starobinsky analysis of the vacuum decay\cite{Starobinsky1994,Starobinsky1996,Starobinsky1997,Starobinsky2023} is needed. Especially  the revolutionary stochastic inflation approach pioneered by Starobinsky\cite{Starobinsky1982,Starobinsky1986}  is extremely useful, although  it requires some modifications.\cite{Pattison2019,Cruces2022,Cruces2022b,Shellard2024} This also includes the so-called separate universe approach, which is somewhat similar to the multi-metric gravity discussed in Sec. \ref{HeatTransferSec}. Here we consider the simple phenomenological scenario based on the energy exchange between the thermal de Sitter vacuum and the created thermal matter. This phenomenological description does not depend on the details of the microscopic (UV) theory, and requires only the slow-roll condition, i.e. the slow variation of the Hubble parameter, $|\dot H| \ll H^2$, or which is the same $|\dot T| \ll T^2$.

The thermal exchange between the de Sitter heat bath and the excited matter generates  the thermal relativistic gas. The temperature of relativistic gas tends to approach  the temperature $T=H/\pi$ of the de Sitter background. Correspondingly, the energy density of this matter  $\epsilon_M$  tends to approach the (quasi)equilibrium value at this temperature, $\epsilon_M(T) \sim T^4$. In terms of the Hubble parameter, one has $\epsilon_M(H) \rightarrow bH^4$, where  the dimensionless parameter $b$ depends on the number of massless relativistic fields. For example, $b=7N_F/120\pi^2$ for $N_F$ species of massless Weyl fermions.

The energy exchange between the vacuum heat bath  and matter can be described by the following dynamical modification of the Friedmann equations,\cite{Volovik2020c}  where the dissipative Hubble friction equation $\partial_t \epsilon_M= - 4H \epsilon_M$ is extended to
\begin{eqnarray}
\partial_t \epsilon_M= - 4H (\epsilon_M - b H^4)\,.
\label{MatterNonConservation}
\end{eqnarray}
This equation describes the tendency of matter to approach the local temperature of the vacuum, $T=H/\pi$. The extra gain of the matter energy, $4bH^5$, is compensated by the corresponding loss of the vacuum energy:
\begin{eqnarray}
\partial_t \epsilon_{\rm vac}= - 4bH^5 \,,
\label{VacuumNonConservation}
\end{eqnarray}

Since the vacuum energy density is $\epsilon_{\rm vac} =6KH^2$, one obtains from Eq.(\ref{VacuumNonConservation}) the following time dependence of the Hubble parameter and of the energy densities of vacuum and matter:
 \begin{eqnarray}
H(t) =b^{-1/3} M_{\rm Pl} \left( \frac{t_{\rm Pl}}{t+t_0}\right)^{1/3}  \,,
\label{DecayLawH}
\\
\epsilon_{\rm vac}(t) =6b^{-2/3} M_{\rm Pl}^4  \left( \frac{t_{\rm Pl}}{t+t_0}\right)^{2/3} \,,
\label{DecayLawV}
\\
\epsilon_M(t) =bH^4 =b^{-1/3}  M_{\rm Pl}^4  \left( \frac{t_{\rm Pl}}{t+t_0}\right)^{4/3}  \,.
\label{DecayLawM}
\end{eqnarray}
Here $M_{\rm Pl}$ is the Planck mass, $M_{\rm Pl}^2 =K$,  and $t_{\rm Pl}=1/M_{\rm Pl}$ is Planck time. We assume that $t_0 \gg t_{\rm Pl}$, and thus $|\dot H| \ll H^2$.

Thus the thermal character of the de Sitter state determines the process of its decay. 
 The obtained power law decay of $H$ in Eq.(\ref{DecayLawH}) was also found in Refs. \cite{Padmanabhan2003,Padmanabhan2005,Markkanen2018,Markkanen2018a,Roman2020,Gong2021}, although using different approaches. 
 In the Padmanabhan model,\cite{Padmanabhan2003,Padmanabhan2005} the de Sitter horizon is considered as the photosphere with the Gibbons-Hawking temperature and with the radiative luminosity $dE/dt \propto T^4 A_H$, where $A_H=4\pi/H^2$ is the area of horizon. Since the energy of the Hubble volume is $E \sim M_{\rm Pl}^2/H$, and 
 $T^4 A_H\sim H^2$, this leads to the power law  for the vacuum energy density in Eq.(\ref{DecayLawV}). As was mentioned by Padmanabhan,\cite{Padmanabhan2005} in his model the late time cosmological constant is independent of its initial value, see Eq.(\ref{DecayLawV}) at $t\gg t_0$.
 
 In  Refs. \cite{Markkanen2018,Markkanen2018a,Roman2020} the Starobinsky stochastic inflation approach has been used. The parameter $b$ therein is proportional to the number $N$ of conformal fields and the parameter $t_0$ is related to the initial value of the Hubble parameter at the beginning of inflation at $t=0$:
  \begin{equation}
 H(t=0) =b^{-1/3}  M_{\rm Pl} \left(\frac{t_{\rm Pl}}{t_0}\right)^{1/3} \ll M_{\rm Pl}
\,.
\label{H0}
\end{equation}
This $H(t=0)$ corresponds to the scaleron mass $M$ in Starobinsky inflation.
The time $t_0\sim E_{\rm Pl}^2/ H^3_{t=0}$ is called the quantum breaking
time of space-times with positive cosmological constant.\cite{Dvali2019,Dvali2022}

All this demonstrates that the phenomenological scenario of thermalization of matter by the de Sitter heat bath in Eqs. (\ref{MatterNonConservation}) and (\ref{VacuumNonConservation}) is rather natural.  It produces the inflation in terms of two phenomenological parameters, $b$ and $t_0$, which determine the decay of the vacuum energy density in Eq.(\ref{DecayLawV}). However, in Sec. \ref{DZeldovichSec} we consider another phenomenological approach, which gives a different power law for the de Sitter decay.

\subsection{Connection to holographic principle}
\label{HolographySec}

The evolution of the gravitational (dark) energy in Eq. (\ref{DecayLawV}) and of the energy of the relativistic
matter in Eq.(\ref{DecayLawM}) allows us to consider another proposal made by Padmanabhan \cite{Padmanabhan2012} (see also Refs.\cite{Krishna2022,Prasanthan2024} and references therein).
It is the holographic postulate which connects the expansion of the Hubble volume with the difference between the number $N_{\rm hor}$ of microstates on the surface of horizon (one degree of freedom per Planck area) and the number $N_{\rm bulk}$ of the degrees of freedom in bulk:
 \begin{eqnarray}
\frac{dV_H}{dt}= G \,(N_{\rm hor}-N_{\rm bulk})\,.
\label{PaddyConjecture2}
\end{eqnarray}
This postulate suggests that the expansion of the universe is being driven towards the holographic
equipartition, so that in the equilibrium de Sitter state all the bulk degrees of freedom inside the horizon can be expressed via the horizon degrees of freedom, $N_{\rm bulk} =N_{\rm hor}$.

Instead of the speculative horizon degrees of freedom, we consider here the corresponding thermodynamic entropy. We already found that according to Eq.(\ref{dSEntropy}) the gravitational entropy of the Hubble volume is equal to the entropy attributed to the cosmological horizon, $S_{\rm bulk}=S_{\rm hor}$. Let us consider what happens in the presence of matter. On one hand, matter adds its contribution $S_M$ to the bulk entropy, and thus $S_{\rm bulk}=S_M+S_{\rm hor}$. On the other hand, matter violates the de Sitter symmetry, which leads to the time dependence of the Hubble volume.

The entropy density of matter  $s_M(t)$ can be obtained from the matter energy density in Eq.(\ref{DecayLawM}):
 \begin{eqnarray}
s_M(t) =\frac{4\pi}{3}bH^3(t) \,.
\label{MatterEntropy1}
\end{eqnarray}
 Then the total entropy of matter $S_M$ in the Hubble volume is time independent:
 \begin{eqnarray}
S_M=s_M(t) V_H(t)= \left(\frac{4\pi}{3} \right)^2 b \,.
\label{MatterEntropy2}
\end{eqnarray}
The time derivative of the Hubble volume $dV_H/dt$, which is obtained from Eq.(\ref{DecayLawH}), is also time independent:
 \begin{eqnarray}
\frac{dV_H}{dt}= \frac{4\pi}{3} b l_{\rm Pl}^2\,.
\label{HubbleDerivative}
\end{eqnarray}
Comparing Eq.(\ref{HubbleDerivative}) with Eq.(\ref{MatterEntropy2}) one has
 \begin{eqnarray}
\frac{dV_H}{dt}= \frac{3}{4\pi}  l_{\rm Pl}^2 S_M\,.
\label{PaddyConjecture3}
\end{eqnarray}
Then, using equation $S_M=S_{\rm bulk} -S_{\rm hor}$, one obtains the general relation between the expansion of the Hubble volume and the difference between the bulk entropy and and its holographic surface value:
 \begin{eqnarray}
\frac{dV_H}{dt}=12G \,(S_{\rm bulk}-S_{\rm hor})\,.
\label{PaddyConjecture1}
\end{eqnarray}
Up to the sign and numerical coefficient, this coincides with Eq.(\ref{PaddyConjecture2}), which supports
the Padmanabhan holographic conjecture,\cite{Padmanabhan2012} but without using the speculative degrees of freedom of the horizon.

\subsection{de Sitter decay and Zel'dovich stiff matter}
\label{DZeldovichSec}

As was mentioned by Padmanabhan, his photosphere model\cite{Padmanabhan2005}  leads to a late time cosmological constant in Eq.(\ref{DecayLawV}), which is independent of the initial value, but its value is still far too large. Can we fix this? In Sec. \ref{GibbsExtensionSec} we obtained indication that the thermodynamics of de Sitter thermal bath has also the properties of the Zel'dovich stiff matter with $w=1$. Let us try the stiff matter scenario using our phenomenological approach. 

We assume that the dynamics of the decaying de Sitter state can be considered as a kind of two-fluid hydrodynamics of superfluid liquid.  We have the (superfluid) vacuum component with $w=-1$, which has de Sitter symmetry, and the (normal) stiff matter component with $w=1$, which violates this symmetry.  These two components tend to approach the common temperature. Such two-fluid behaviour of the de Sitter state may also come from the observation in Sec. \ref{ThermalFluctuationsSec} that thermal fluctuations of the energy density in the Hubble volume are on the order of the vacuum energy density itself, $<(\Delta \epsilon_{\rm vac})^2> / <\epsilon_{\rm vac}>^2 \sim 1$.
So, in these speculations, the de Sitter state behaves as a mixture of dark energy (the de Sitter vacuum) and dark matter (the stiff matter or the thermal fluctuations of de Sitter).

For the (dark) matter with $w=1$, the dissipative Hubble friction equation is $\partial_t \epsilon_{\rm DM}= - 6H \epsilon_{\rm DM}$.  Due to the energy exchange between the gravitational (dark energy) component and the dark matter component,  the temperature of dark matter tends to approach the heat bath temperature $T=H/\pi$. Then instead of Eqs.(\ref{MatterNonConservation}) and (\ref{VacuumNonConservation}) one obtains
\begin{eqnarray}
\partial_t \epsilon_{\rm DM}= - 6H (\epsilon_{\rm DM}- \tilde b H^2)\,,
\label{StiffNonConservation}
\end{eqnarray}
and
\begin{eqnarray}
\partial_t \epsilon_{\rm vac}= - 6\tilde bH^3 \,.
\label{VacuumNonConservationS}
\end{eqnarray}
Here $\tilde b$ is the phenomenological dimensionless parameter on the order of unity, which microscopic origin is to be found.
The similar equations with the corresponding phenomenological dimensionless parameter  $\gamma$ were suggested in Ref.\cite{Klinkhamer2012}, where  the Polyakov scenario\cite{Polyakov2012} of the infrared instability of the de Sitter space was discussed and the pressureless matter was considered with $w=0$.  

The Eq.(\ref{VacuumNonConservationS}) gives the following power law decay of dark energy and dark matter:
\begin{eqnarray}
\epsilon_{\rm vac}(t) \sim \epsilon_{\rm DM}(t)   \sim M_{\rm Pl}^4  \left( \frac{t_{\rm Pl}}{t+t_0}\right)^2\,.
\label{DecayLawS}
\end{eqnarray}
For large $t\gg t_0$ this gives the reasonable order of magnitude of the vacuum energy density and of the energy density of matter in the present time:
\begin{eqnarray}
\epsilon_{\rm vac}(t_{\rm present})\sim \epsilon_{\rm DM}(t_{\rm present})\sim\frac{M_{\rm Pl}^2}{t_{\rm present}^2} \sim 10^{-120} M_{\rm Pl}^4\,.
\label{VacuumPresent}
\end{eqnarray}

The same behaviour of the vacuum energy density was obtained using the Hawking 4-form field.\cite{KlinkhamerVolovik2008c,Volovik2013}. In this case the role of dark matter is played by the oscillations of the 4-form field during decay.\cite{KlinkhamerVolovik2017} Note that in our approach, both dark energy and dark matter come from the gravitational degrees of freedom. In this sense it has relations to Refs. \cite{Cano2024a,Cano2024b,DiFilippo2024} and references therein, where the role of the gravitational degrees of freedom is discussed.

\subsection{Thermal fluctuations in de Sitter state}
\label{ThermalFluctuationsSec}

Here we consider thermal fluctuations in the de Sitter thermal state, which also may serve as the source of the two-fluid behaviour of the de Sitter dynamics. The de Sitter thermal state represents the excited state of the Minkowski quantum vacuum. In addition, the deep Minkowski quantum vacuum experiences the thermal fluctuations, which may play the role of the dark matter.  

According to Landau-Lifshitz,\cite{LandauLifshitz} the thermal fluctuations are determined by the compressibility of the system and the considered volume $V$. In case of the fluctuating relativistic vacuum one has:\cite{Volovik2013}
\begin{equation}
\left<(\Delta \epsilon_{\rm vac})^2\right> =\left<(\Delta P_{\rm vac})^2\right> =\frac{T}{V\chi_{\rm vac}} \,.
\label{ThermalFluctuations}
\end{equation}
Here $\chi_{\rm vac}$ is the vacuum compressibility\cite{KlinkhamerVolovik2008} -- the compressibility of the fully equilibrium Minkowski vacuum  with  $\epsilon_{\rm vac}=-P_{\rm vac}=0$.  

Note the main difference between the thermal fluctuations and quantum fluctuations. The contribution of the quantum fluctuations of the relativistic quantum fields to the vacuum energy density is typically on the order of $M_{\rm Pl}^4$, where $M_{\rm Pl}$ is the Planck mass. But in the fully equilibrium vacuum state this contribution is cancelled by the ultraviolet trans-Planckian degrees of freedom due to the thermodynamic Gibbs-Duhem relation.\cite{KlinkhamerVolovik2008,Volovik2013} This cancellation is universal, being valid both for the relativistic vacuum states and for the non-relativistic grounds states of the condensed matter systems. But the contribution of thermal fluctuations to vacuum energy is in the range of applicability of infrared physics, where it is expressed in terms of the temperature $T$ and the compressibility of the vacuum.

The value of the vacuum compressibility is determined by the ultraviolet physics\cite{KlinkhamerVolovik2008} 
with its Planck energy scale, $\chi_{\rm vac}^{-1}\sim M_{\rm Pl}^4$. This is similar to the gravitational coupling, which is also determined by the Planck scale, $K\sim M_{\rm Pl}^2$. On the other hand, the temperature corrections to 
$\chi_{\rm vac}^{-1}$ and $K$ as well as the Casimir corrections are within the range of applicability of the infrared physics, see Ref. \cite{VolovikZelnikov2003} for the universal temperature correction to the gravitational coupling $K$. 

The contribution to the compressibility from the infrared physics  was discussed in Refs.\cite{BLHu2022,BLHu2024,BLHu2024a,Widom1998}. The negative contributions to compressibility obtained in these papers do not violate the stability of the quantum vacuum, since these contributions represent the corrections, which are small compared to the main value of the vacuum compressibility,  $\chi_{\rm vac}^{-1}\sim M_{\rm Pl}^4$. 

Taking into account that the excited vacuum (the de Sitter state) has the temperature $T=H/\pi$  and the energy density  $<\epsilon_{\rm vac}> \sim M_{\rm Pl}^2H^2$, it follows from Eq.(\ref{ThermalFluctuations})  that the thermal fluctuations of the energy density in the Hubble volume are on the order of thermal energy energy density:
\begin{equation}
\left<(\Delta \epsilon_{\rm vac})^2\right>_{V=V_H} \sim \left<\epsilon_{\rm vac}\right>^2    \,.
\label{ThermalFluctuations2}
\end{equation}
 This can be the reason of the two-fluid behaviour of the de Sitter state discussed in 
 Sec. \ref{DZeldovichSec}. The thermal fluctuations of dark energy play the role of dark matter in the same way as the oscillations of the dark energy in Ref.\cite{KlinkhamerVolovik2017}, and they lead to the same power law decay in Eq.(\ref{DecayLawS}) and to the present values of dark energy and dark matter in Eq.(\ref{VacuumPresent}).

\subsection{Cosmological constant problems}
\label{CCProblemsSec}

In principle, the phenomenological approaches to the dynamics of the vacuum energy density may produce different power-law decays. Examples are  Eq.(\ref{DecayLawS}) and Eq.(\ref{DecayLawV}). However, it is not excluded that these two asymptotic laws may correspond to different epochs.

So, if this speculative approach in Sec. \ref{DZeldovichSec} works, the equation (\ref{VacuumPresent}) may solve all the cosmological constant problems:

1) why the cosmological constant is not large;

2) why the dark energy is on the order of magnitude of dark matter;

3) why they have the present value.

\section{From de Sitter to black hole thermodynamics}
\label{BHsec}

\subsection{de Sitter vs black hole} 
\label{dSvsBH}

The thermodynamics of de Sitter state is very different from the thermodynamics of black holes. Black hole is the compact object. The temperature of the Hawking radiation $T_H$ from the black hole horizon is well determined, which is also supported by the condensed matter analogs.\cite{Unruh1981}  On the other hand, the origin of the black hole entropy is still not clear, although it can be determined from the equation $dM = T_H dS$, assuming that the laws of thermodynamics are applicable to this compact object. 
 
On the contrary, the de Sitter state is not the compact object. It is the homogeneous vacuum state without boundaries that has the homogeneous energy density as the local thermodynamic variable. The local energy density allows us to also introduce the local temperature $T=H/\pi$. This local temperature can be measured by any detector which is stationary with respect to the shift velocity and which measures for example the rate of the ionization of an atom, $w\propto \exp(-E/T)$.

\subsection{Entropy of expanding, contracting and static de Sitter}
\label{ContractingDSSec}

However, the de Sitter state allows us to probe the origin of the black hole entropy. For that we must consider three different states of the de Sitter quantum vacuum: expanding de Sitter with $H>0$, contracting de Sitter with $H<0$ and the de Sitter state with the fully static metric. Let us note that the metrics of the expanding and contracting de Sitter states are stationary, but not static, since their shift velocities are non-zero. 

Let us start with the fully static de Sitter metric:
 \begin{equation}
ds^2= - (1-H^2r^2)dt^2 +   \frac{dr^2}{1-H^2r^2}+r^2 d\Omega^2\,.
\label{static}
\end{equation}
This metric has singularity at the cosmological horizon, which requires the proper attention. In one approach this static configuration can be considered as the intermediate symmetric state between the two states with broken time reversal symmetry -- the expanding and contracting states. The same consideration applies to the static hole, that can be viewed as the intermediate state between the black and white holes.\cite{Volovik2022}.   

The contracting de Sitter vacuum has negative Hubble parameter, $H<0$. That is why its local temperature is negative, $T=H/\pi <0$, and the local entropy is also negative, $s_{\rm contracting}=3H/4G=-3/(4Gr_H)<0$. The total entropy in the Hubble volume of contracting de Sitter is 
\begin{equation}
 S_{\rm contracting}=s_{\rm contracting}V_H=-\frac{A}{4G}
\,,
\label{EntropyContracting}
\end{equation}
where $A$ is the area of the cosmological horizon. 
That is why the entropy of the static de  Sitter in Eq.(\ref{static}) as the intermediate state between the states with positive and negative entropies is zero. Correspondingly the temperature of this state must be infinite, which is consistent with the singularity at the horizon, see also Ref.\cite{Milekhin2024}. 

\subsection{Gravastar -- black hole with de Sitter core}
\label{GravastarSec}

The connection between the black hole and de Sitter appears, when we consider the black hole obtained by the deformation of the gravastar.\cite{Volovik2023a} The gravastar is an object, which contains the de Sitter vacuum  inside the black hole horizon.\cite{Chapline2003,Mazur2023,Mottola2023} We consider the gravastar in which the black hole horizon coincides with the de Sitter horizon, $r_{\rm bh}=r_H$ (the metric in the state with the critical value of the mass parameter $m=2MG|H|=1$ at which two horizons merge is illustrated in Fig. 1 of Ref. \cite{Dymnikova2002}).  Since in such gravastar the two horizons cancel each other, there is no Hawking radiation and the entropy of the gravastar is zero. In the Painlev\'e-Gullstrand form the metric of such gravastar is given by Eq.(\ref{PG1}) with the following shift velocity:
\begin{eqnarray}
v(r)= -\sqrt{\frac{r_H}{r}} \,\,,\,\, r >r_H \,,
\label{v1}
\\
v(r)=- \frac{r}{r_H} \,\,,\,\, r < r_H \,.
\label{v2}
\end{eqnarray}
Here $r_H=1/|H|=r_{\rm bh}$, where $r_{\rm bh}=2MG$ and $M=1/(2G|H|)$ is the mass of the black hole, which is formed by the de Sitter core.

The shift  velocity $v(r)$ is continuous across the surface $r=r_H$, while the gradient of the shift velocity $dv/dr$ experiences jump at this surface. It is important that the shift velocity $v(r)$ is negative everywhere.  Since it is negative in the de Sitter core, this means that the de Sitter spacetime in the core of this gravastar is contracting, $v(r)=Hr=-r/r_H<0$, i.e. the Hubble parameter is negative, $H=-1/r_H<0$. 

\subsection{Entropy of black hole from negative entropy of contracting de Sitter}
\label{BHentropySec}

According to Eq.(\ref{EntropyContracting}) the  region of the contracting de Sitter core of the gravastar has negative entropy, $S_{\rm contracting}=-A/4G$. That is why the gravastar is unstable towards the shrinking of the de Sitter region, since this leads to the increase of the entropy of the whole system due to decrease of the negative entropy of the  contracting de Sitter state. Due to the energy conservation the shrinking of the volume of the de Sitter core leads to formation of the singularity ar $r=0$, where the mass becomes concentrated. In the final state -- the black hole -- the de Sitter region with negative entropy fully disappears by shrinking to the singularity with mass $M$. The resulting black hole with mass $M$ acquires the positive entropy, $A/4G$:
 \begin{equation}
 S_{\rm BH}= S_{\rm gstar} - S_{\rm contracting}=0 -  s_{\rm contracting}V_H = \frac{A}{4G} \,.
\label{BHentropy}
\end{equation}
The black hole entropy is also concentrated in the singularity, together with the curvature ${\cal R}$ and mass $M$.\cite{Volovik2023a} This again supports the holographic connection between the entropy of bulk, which is concentrated in the singularity, and the surface entropy of the black hole horizon.

Eq.(\ref{BHentropy}) allows to interpret the zero value of the entropy of the gravastar in terms of the cancellation of entropies of two horizons. In the initial gravastar state the entropy of the contracting de Sitter horizon fully compensates the entropy of the  black hole horizon:
 \begin{equation}
 S_{\rm gstar}=S_{\rm contracting}+S_{\rm BH}=-\frac{A}{4G}+\frac{A}{4G} =0\,.
\label{Gravastar}
\end{equation}

\subsection{White hole and anti-gravastar}
\label{antigravastarSec}

In the same way the anti-gravastar can be obtained as the white hole with the de Sitter core. This object  has  the following shift velocities:
\begin{eqnarray}
v(r)= \sqrt{\frac{r_H}{r}} \,\,,\,\, r >r_H \,,
\label{antiv1}
\\
v(r)= \frac{r}{r_H} \,\,,\,\, r < r_H \,.
\label{antiv2}
\end{eqnarray}
It is obtained from the pure white hole, which has the negative horizon entropy $S_{\rm WH}=-A/4G$,\cite{Volovik2022,Volovik2020} 
by growing the de Sitter state in its core with positive entropy. In the anti-gravastar state the entropies of two horizons cancel each other in the same way as in the gravastar:
 \begin{equation}
 S_{\rm antigstar}=S_{\rm expanding}+S_{\rm WH}=\frac{A}{4G}-\frac{A}{4G} =0\,.
\label{antiGravastar}
\end{equation}

It is important that the coordinate singularities in the metric of the gravastar and in the metric of the anti-gravastar can be smoothly removed by small deformations. That is why these objects do not depend on the choice of the coordinate systems, and thus they are equivalent to the fully static black hole with the fully static de Sitter core:
\begin{eqnarray}
ds^2= - (1-2GM/r)dt^2 +   \frac{dr^2}{1-2GM/r}+r^2 d\Omega^2 \,\,, \,\, r>2MG=1/H\,,
\label{staticG1}
\\
ds^2= - (1-H^2r^2)dt^2 +   \frac{dr^2}{1-H^2r^2}+r^2 d\Omega^2 \,\,, \,\, r<2MG=1/H\,.
\label{staticG2}
\end{eqnarray}
This again supports the zero entropy of these gravastars, and also the zero entropy of the fully static de Sitter with metric in Eq.(\ref{static}).

\subsection{Gibbs-Duhem and black hole thermodynamics}
\label{AgainBlackSec}

Let us show, that the modified Gibbs-Duhem relation in Eq.(\ref{GibbsDuhem}) is applicable also to the thermodynamics of black holes.
As distinct from the de Sitter state, the black hole is the compact object, and its thermodynamics is operating with the global parameters, such as mass $M$, entropy of horizon, total electric charge $Q$ and total angular momentum $J$. This global thermodynamics can be described by the integral form of the  Gibbs-Duhem relation in Eq.(\ref{GibbsDuhem}). The curvature here comes from the central singularity:\cite{Balasin1993} 
\begin{equation}
{\cal R}=8\pi MG\,\delta({\bf r})\,.
\label{CentralSingularity}
\end{equation}
Since the energy density here is $\epsilon=M\delta({\bf r})$,   the integration of the right-hand-side of Eq.(\ref{GibbsDuhem}) over space gives the following relation for the Schwarzschild black hole: 
\begin{equation}
T_{\rm BH} S_{\rm BH} =M-\int d^3 r \sqrt{-g} K{\cal R}= \frac{M}{2}\,.
\label{BlackGibbs}
\end{equation}
This agrees with global thermodynamics  of the black hole with the Hawking temperature $T_{\rm BH}=\frac{1}{8\pi MG}$ and the Bekenstein-Hawking entropy $S_{\rm BH}=A/4G$. The Eq.(\ref{BlackGibbs}) is valid also for the white hole, where temperature and entropy are opposite to that of the black hole with the same mass,\cite{Volovik2022} 
 $T_{\rm WH}(M)=- T_{\rm BH}(M)$ and  $S_{\rm WH}(M)=- S_{\rm BH}(M)$.
 
In principle, the central singularity in the black hole may spontaneously loose the spherical symmetry, forming for example a kind of rigid top. This in turn may influence the shape of the event horizon.
This is similar to the deformations of the cosmological horizon by masses concentrated on vertices of Platonic solids deep within the Hubble volume.\cite{Fischler2024}

\subsection{Entropy of the Schwarzschild-de Sitter cosmological horizon}
\label{SdSsection}

Let us consider the possible application of the modified Gibbs-Duhem relation to the Schwarzschild-de Sitter (SdS) black hole. We discuss the simple case of the Nariai limit, when the black hole horizon approaches the cosmological horizon, $r_b \rightarrow r_0-0$ and  $r_c \rightarrow r_0 + 0$, where
\begin{equation}
r_0=(GM/H^2)^{1/3} =\frac{1}{\sqrt{3}H} \,.
\label{r0}
\end{equation}
 In this limit the temperatures of the horizons approach the Bousso-Hawking value:\cite{BoussoHawking1996} 
\begin{equation}
T_b=T_c=\frac{\sqrt{3}}{2\pi}H=\frac{1}{6\pi GM}\,.
\label{GlobalT}
\end{equation}

 The entropy of the cosmological horizon $S_{\rm c} $ can be obtained by integration of the right-hand-side of Eq.(\ref{GibbsDuhem}) over the vollume inside the cosmological horizon with radius $r_0$:
\begin{equation}
T_{\rm c} S_{\rm c} =M-\int _{r<r_0}d^3 r \sqrt{-g} K{\cal R}  \,.
\label{SdSGibbs}
\end{equation}
Since we discuss here the Nariai limit, one obtains:
\begin{equation}
T_{\rm c} S_{\rm c} =M-\int_{r<r_0} d^3 r \sqrt{-g} K{\cal R}  = M- \frac{M}{2} + \frac{3 H^2}{4\pi} \, \frac{4\pi r_0^3}{3}= \frac{3}{2}M.
\label{SdSGibbs2}
\end{equation}
From this equation one obtains the entropy of the cosmological horizon:
\begin{equation}
S_{\rm c}=\frac{3M}{2T_{\rm c}} =\frac{\pi r_0^2}{G}= \frac{A}{4G}\,.
\label{SdSentropy}
\end{equation}
This again agrees with the Gibbons-Hawking entropy attributed to the horizon, although this holographic connection is valid only in the Narai limit.

\subsection{Heat exchange between black holes in the multi-metric ensemble}
\label{ExchangeBetweenBHs}

The thermodynamic character of the black hole objects can be tested using the multi metric ensemble in Sec. 
\ref{HeatTransferSec}. Let us consider ensemble of $n$ sub-universes in the same coordinate spacetime,
$S=-\int d^4 x \sum_{n} {\cal L}_n$, but with with different metrics $g_{\mu\nu(n)}$, different gravitational couplings $K_n$, and the corresponding individual black holes with masses $M_n$. If one introduces the energy exchange between these sub-universes, then the masses $M_n$ will be varied at the fixed total mass $M=\sum_n M_n$ of the whole Universe. In the thermal equilibrium, which corresponds to the maximum of the total entropy, the individual metrics approach the common metric, and the whole system corresponds to the Universe  with the  gravitational coupling $K=\sum_n K_n$ and with the black hole of mass $M$. The equilibrium masses $M_n$ of the individual substates approach the values:
 \begin{equation}
M_n=M\,\frac{K_n}{K}    \,.
\label{MultiBH}
\end{equation}
In this equilibrium states, the Hawking temperature of the black hole with mass $M$ in the whole ensemble coincides with Hawking temperatures of the individual black holes in each gravitational sub-states:
\begin{equation}
\frac{1}{T_n}=  \frac{dS_n}{dM_n}=\frac{dS}{dM}=\frac{1}{T_H} \,.
\label{MultiBH2}
\end{equation}
This is the effect of thermalization.

\section{Black and white holes entropy from macroscopic quantum tunneling }
\label{macroTunnel}

\subsection{Collective canonically conjugate variables for Schwarzschild black hole}
\label{CollectiveVar}

Considering the thermodynamics of gravity systems we used the thermodynamically conjugate variables --
the gravitational coupling $K=1/16\pi G$ and the scalar curvature ${\cal R}$, which as the covariant quantity may serve as one of the thermodynamical characteristics of the macroscopic matter.\cite{Pronin1987}  
 These variables are the local thermodynamic variables, which are similar to temperature, pressure, chemical potential, number density, etc., in condensed matter physics.   

The gravitational coupling $K$ is determined by the UV microscopic physics, but in the description of gravity as macroscopic phenomenon in the IR limit it is the collective variable.
In the  superfluid $^3$He-A with Weyl fermionic quasiparticles, both the coupling $K$ in the effective gravity and  the fine structure  ``constant`` $\alpha$ in the effective electrodynamics are determined by physics on the microscopic (atomic) level.\cite{Volovik2003} In the microscopic theory, one obtains $K\propto \Delta_0^2$ and $1/\alpha \propto \ln (\Delta_0/T)$, where $\Delta_0$ is the gap amplitude  and $T$ is the temperature of the liquid.  The gap amplitude $\Delta_0$  plays the role of the ultraviolet cut-off $M_{\rm Pl}$, while $T$ provides the infrared cut-off.  For the relativistic quantum fields with massless particles, the infrared cut-off is either the temperature $T$ or the strength of the fields. In the inhomogeneous superfluid (inhomogeneous vacuum) both $K$ and $\alpha$ depend on coordinates. It is not surprizing that in the relativistic quantum vacuum there is also the connection  between the gravitational coupling $K$ and the coupling $\alpha$  in quantum electrodynamics as suggested in Refs. \cite{Landau1955,Akama,AkamaTerazawa,Terazawa,TerazawaAkama,Terazawa1981,KlinkhamerVolovik2005}

Now, for the discussion of the quantum-mechanical tunneling of the macroscopic objects we need the collective dynamical variables, instead of the thermodynamic variables.
The canonically conjugate dynamical variables, which are relevant for the black hole, are the gravitational coupling $K$ and the horizon area $A=4\pi R^2$. Bekenstein \cite{Bekenstein1974}  proposed that $A$ is an adiabatic invariant and thus can be quantized according to the Ehrenfest principle, that classical adiabatic invariants may
correspond to observables with discrete spectrum. So the area of the horizon $A$ is the proper candidate for the quantum mechanics of the black hole. 
 
\subsection{Modified first law of black hole thermodynamics }
\label{FirstLaw}

Let us first consider how these variables enter the black hole thermodynamics. For that it is convenient to use the redefined gravitational coupling $\tilde K =4\pi K =1/4G$. In terms of this coupling 
the Hawking temperature of Schwarzschild black hole and its Bekenstein entropy are:
 \begin{equation}
  T_\text{BH}=\frac{\tilde K}{2\pi M}~~,~~S_\text{BH}
             = \frac{\pi M^2}{\tilde K}=A\tilde K   \,.
\label{eq:HawkingT}
\end{equation}
 If $\tilde K$ is a global thermodynamic variable, one obtains the following modification of the first law of black hole thermodynamics:
\begin{eqnarray}
dS_\text{BH}=d(A\tilde K)= \pi d (M^2/\tilde K)  = - \pi \frac{M^2}{\tilde K^2}d\tilde K + 2\pi \frac{M}{\tilde K} dM
\label{dS2}
\end{eqnarray} 
or
 \begin{eqnarray}  
 dS_\text{BH}= -Ad\tilde K + \frac{dM}{T_\text{BH}} \,.
\label{FirstLawEq}
\end{eqnarray} 
This modification is similar to the modification in terms of the moduli fields.\cite{Gibbons1996} But in our case  the thermodynamic variable, which is  conjugate to the thermodynamic variable $\tilde K$, is the product of the black hole area and the black hole temperature, $AT_\text{BH}$. On the other hand in dynamics, $\tilde K$ and $A$ are canonically conjugate variables, see Sec.\ref{Canonical}. 

In general, the variable $\tilde K$ is local and depends on space coordinate, but in the same way as for the moduli fields,\cite{Gibbons1996} the black hole thermodynamics is determined by the asymptotic value of $K$ at spatial infinity. In Eq.(\ref{FirstLawEq}), $\tilde K\equiv \tilde K({\infty})$ is the global quantity, which characterizes the quantum vacuum in full equilibrium, i.e. far from the black hole. 

Also, the variable $\tilde K$  allows us to study the transition to the vacuum without gravity, i.e. to the vacuum where $\tilde K \rightarrow\infty$ and thus $G\rightarrow 0$, see Sec.\ref{Canonical}.

\subsection{Adiabatic change of coupling $K$ and adiabatic invariant}
\label{adiabatic}

Let us change the coupling $\tilde K$ and the black hole mass $M$ adiabatically, i.e. at constant entropy of the black hole. Then the equation $dS_\text{BH}=0$ gives 
\begin{equation}
\frac{dM}{d\tilde K}= AT_\text{BH}=\frac{M}{2\tilde K} \,.
\label{ratio}
\end{equation}
This shows that $M^2/\tilde K={\rm const}$ is the adiabatic invariant for the spherical electrically neutral black hole. Thus according to  the Bekenstein conjecture \cite{Bekenstein1974}, it can be quantized in quantum mechanics:
\begin{equation}
\frac{M^2}{\tilde K}= a N \,.
\label{quantization}
\end{equation}
Here $N$ is integer, and $a$ is some fundamental dimensionless parameter of order unity.
If this conjecture is correct, one obtains the quantization of the entropy of Schwartzschild black hole:
\begin{equation}
S_\text{BH}(N)=\pi \frac{M^2}{\tilde K}= \pi a N \,.
\label{quantizationSchwarz}
\end{equation}

The Bekenstein idea on the role of adiabatic invariants in quantization of the black hole requires further consideration, see some approaches to that in Refs. \cite{Horowitz1996,Barvinsky2001,Barvinsky2002,Ansorg2012,Visser2012,Tharanath2013}.  In particular, the similarity between the energy levels of Schwarzschild black hole and the hydrogen atom has been suggested. \cite{Bekenstein1997,Corda2015}
We leave this problem for the future. This consideration should be supported by microscopic theory, see e.g., Ref. \cite{Carlip2014}. 

In this respect, the condensed matter analogs can be useful, since in the condensed matter systems the physics is known both on macro and micro (atomic) levels. One example is provided by consideration of the quantum nucleation of the vortex ring in moving superfluids --  the vortex instanton. It shows that the corresponding entropy which determines the nucleation process, $\exp(-S_\text{ring})$,  is quantized:
\begin{equation}
S_\text{ring}(N)=2\pi N \,.
\label{quantizationSchwarz}
\end{equation}
Here $N$ is the number of atoms involved in the process of the vortex instanton, see Sec. 26.4 and Eqs.(26.20)-(26.21) in Ref. \cite{Volovik2003}. This would correspond to the parameter $a=2$ in Eq.(\ref{quantization}).

\subsection{$A$ and $K$ as canonically conjugate variables and black-hole -- white-hole quantum tunneling}
\label{Canonical}

The canonically conjugate variables  $A$ and $K$ allow us to consider the quantum mechanical tunneling from the black hole to the white hole -- the process  discussed in Refs. \cite{Barcelo2014,Barcelo2017,Rovelli2018,Rovelli2019,Rovelli2018b,Uzan2020b,Uzan2020c,Uzan2020d,Bodendorfer2019,Rovelli2021a} and in references therein.   In  the semiclassical description of this macroscopic quantum tunneling, the trajectory in the $(\tilde K,A)$ phase space must be considered, which in the complex plane connects the black and white holes.

Note the difference from consideration in Sec. \ref{adiabatic}, where $\tilde K$ varies in the adiabatic regime, i.e. at fixed entropy $S_{\rm BH}$, while the area $A$, temperature $T_{\rm BH}$ and mass $M$ follow the variation of $\tilde K$.  In the dynamic regime, which is relevant for the description of quantum tunnelling, the parameter $\tilde K$ varies at fixed energy (fixed mass $M$ of the black hole), while the area $A$, the temperature $T_{\rm BH}$ and entropy $S_{\rm BH}$ follow the variation of $\tilde K$.

As in the case of the semiclassical consideration of the Hawking radiation in terms of the quantum tunneling,
we shall use the Painleve-Gullstrand coordinate system  with the metric:
 \begin{equation}
ds^2= - dt^2(1-{\bf v}^2) - 2dt\, d{\bf r}\cdot {\bf v} + d{\bf r}^2 \,,
\label{PGmetric}
\end{equation}
where for the Schwartzschild black hole one has
\begin{equation}
{\bf v} ({\bf r})=\mp \hat{\bf r} \sqrt{\frac{R}{r}}=\mp \hat{\bf r} \sqrt{\frac{M}{2r \tilde K}}=\mp \hat{\bf r} \sqrt{\frac{2MG}{r}}\,,
\label{velocity}
\end{equation}
where $R$ is the radius of the horizon; the minus sign corresponds to the black hole and the plus sign describes the white hole.
In the theory with the variable gravitational coupling, the sign changes at the singularity $\tilde K=\infty$ (i.e. at $G=0$), when 
the black hole shrinks to a point and then expands as a white hole. The point $\tilde K=\infty$, where gravity disappears,  serves as the branch point, where the velocity of the freely falling observer changes sign. 

The vector ${\bf v}$ is normal to the surface of the horizon. When ${\bf v}$ changes sign, the horizon area, $A$, also changes sign: it crosses zero at $\tilde K=\infty$ and  becomes negative on the white-hole side of the process, $A \rightarrow -A$. This also could mean that due to connection between the area and entropy the white hole may have negative entropy. This is what we shall discuss in the section \ref{WHentropy}.

The quantum tunneling exponent is usually determined by the imaginary part of the action on the trajectory, which transforms the black hole to white hole. In terms of Euclidean action one has:
 \begin{equation}
p_{\text{BH} \rightarrow \text{WH}}\propto \exp{\left(-I_{\text {BH} \rightarrow \text{WH}}\right)} \,\,,\,\, I_{\text{BH} \rightarrow \text{WH}} =  \int_C A(\tilde K')d\tilde K' \,.
\label{TunnelingExponentGen}
\end{equation}
Here the semiclassical trajectory $C$ is at $M=\rm{const}$, and thus $A(\tilde K')=\pm \pi M^2/(\tilde K')^2$. Along this trajectory the variable $\tilde K'$ changes from $\tilde K$ to the branch point at $\tilde K'=\infty$, and then from  $\tilde K'=\infty$ to $\tilde K'=\tilde K$ along the other  branch, where the area $ A(\tilde K')<0$.  
The integral gives the tunneling exponent of the transition to the white hole 
\begin{equation}
I_{\text{BH} \rightarrow \text{WH}}=  \,  2\pi M^2     \int_{\tilde K}^\infty \frac{d\tilde K'}{\tilde K'^2}= 2 \pi \frac{M^2}{\tilde K}  \,.
\label{TunnelingExponent}
\end{equation}
The tunneling exponent in Eq.(\ref{TunnelingExponent}) can be expressed in terms of the black hole entropy in Eq (\ref{eq:HawkingT}), and for the probability of transition one obtains:
\begin{equation}
p_{\text{BH} \rightarrow \text{WH}}\propto \exp{\left(-2\pi M^2/\tilde K\right)}=\exp{\left(-2S_\text{BH}\right)}
\,.
\label{tunneling}
\end{equation}

It is important that Eq.(\ref{tunneling}) contains exactly twice the black hole entropy.
This allows us to consider again the entropy of the white hole.

\subsection{Negative entropy of white hole}
\label{WHentropy}

The connection between the probability of the transition and the thermodynamic fluctuations\cite{Landau_Lifshitz} is applicable also to the transition between the black and white holes.  The total change of the entropy in this  process is
$\Delta S=S_\text{WH}-S_\text{BH}$. According to Eq.(\ref{tunneling}) this change is equal to $-2S_\text{BH}$. Then, from equation $S_\text{WH}-S_\text{BH}=-2S_\text{BH}$, one obtains that the entropy of the white hole is equal with the opposite sign to the entropy of the black hole with the same mass: 
\begin{equation}
S_\text{WH}(M)=-S_\text{BH}(M)
\,.
\label{fluctuation}
\end{equation}
This means that the white hole, which is obtained by the quantum tunneling from the black hole and thus has the same mass $M$ as the black hole, has the negative temperature 
 $T_\text{WH}=- T_\text{BH}$ and the negative area $A_\text{WH}=- A_\text{BH}$. According to the first law in Eq. (\ref{FirstLawEq}) applied now to the white hole, this gives the negative entropy, $S_\text{WH}=- S_\text{BH}$.
 
 Such anti-symmetry between the black and white holes is similar to the anti-symmetry between the expanding and contracting de Sitter state discussed in Sec. \ref{ContractingDSSec}.
 
 The discussed transition from the black hole to the white hole with the same mass $M$ is not the thermodynamic transition. It is the quantum process of tunneling between the two quantum states. It is one of many routes of the black hole evaporation, including formation of small white hole on the late stage of the decay. \cite{Barcelo2014,Barcelo2017,Rovelli2018,Rovelli2019,Rovelli2018b,Uzan2020b,Uzan2020c,Uzan2020d,Bodendorfer2019,Rovelli2021a}
The uniqueness of this particular route, the hidden information and the (anti)symmetry between the black and white holes are combined to produce the negative entropy of the white hole.

\subsection{Black hole to white hole transition as series of the Hawking radiation co-tunneling}
\label{cotunneling1}

Let us show that the result (\ref{tunneling}), where the probability of the quantum tunneling between the black and white holes transition is determined by the twice the black hole entropy, is supported by consideration of the conventional Hawking radiation of particles from the black hole. 
The tunneling exponent of radiation can be expressed in terms of the change in the entropy of the black hole after radiation,  $p\propto e^{\Delta S_{\text{BH}}}$, see Refs. \cite{Kraus1997,Parentani1997,Berezin1999,Wilczek2000}. This demonstrates that the quantum process of Hawking radiation can be considered as thermodynamic fluctuation.\cite{Landau_Lifshitz}  

We consider the process of the co-tunneling, in which the particle escapes the black hole by quantum tunneling, and then this particle tunnels to the white hole through the white hole horizon (this is the analog of the electron tunneling via an  intermediate virtual state in electronic systems \cite{Feigelman2005,Glazman2005}). This process takes place at the fixed total mass $M$. The tunneling exponent for this process  to occur is $e^{(\Delta S_{\text{BH}}+\Delta S_{\text{WH}})}$. 
Summation of all the processes of the tunneling of matter from the black hole to the formed white hole gives finally Eq.(\ref{tunneling}):
\begin{equation}
p\propto e^{\sum(\Delta S_{\text{BH}}+\Delta S_{\text{WH}})}=e^{2\sum\Delta S_{\text{BH}}}=\exp{\left(-2S_\text{BH}\right)}
\,.
\label{tunneling2}
\end{equation}
Here we took into account the (anti)symmetry in the dynamics of black and white holes in the process of quantum tunneling,
$\sum\Delta S_{\text{BH}}=\sum \Delta S_{\text{WH}}$.

\subsection{Emission of small black holes vs Hawking radiation}
\label{EmissionSmall}

The principle that the macroscopic quantum  tunneling can be considered as thermodynamic fluctuation can be also applied to the process of the creation of  pairs black holes,\cite{HawkingHorovitz1995} to the process of splitting of the black hole into two or several smaller black holes with the same total mass, see e.g. Ref.\cite{HyeyounChung2011}, and to other processes with macroscopic objects. For example, the probability of the decay  of the black hole with mass $M$ into two black holes with $M_1+M_2=M$ is:
\begin{eqnarray}
p(M \rightarrow M_1+M_2) \propto \,
\\
e^{S_{\text{BH}}(M_1)+S_{\text{BH}}(M_2)-S_{\text{BH}}(M_1+M_2)} =e^{-8\pi G M_1M_2}.
\label{tunnelingM1M2}
\end{eqnarray}

In the particular limit case $m=M_1 \ll M_2\approx M$, this channel of the black hole decay describes the emission of the small black hole with mass $m$ by the large black hole with mass $M$:
 \begin{equation}
p(m,M)\propto \exp{\left(-\frac{m}{T_{\rm BH}(M)}\right)}  \,\,,\,\, m \ll M
\,.
\label{tunnelingMm}
\end{equation}
This shows that the macroscopic tunneling process of emission of a small black hole  by a large black hole is governed by the same Hawking temperature 
$T_{\rm BH}(M)=1/(8\pi GM)$ as the Hawking radiation of a particle, which tunnels across the horizon. 

However, there is the difference.
In Eq.(\ref{tunnelingMm}) the quadratic term $m^2$ is neglected. In general case, one obtains from Eq.(\ref{tunnelingM1M2}):
 \begin{equation}
p(m,M-m) \propto \exp{\left(-8\pi Gm(M-m)\right)}
\,.
\label{tunnelingMmquadratic}
\end{equation}
This equation demonstrates the effect of back reaction -- the correction to the Hawking radiation caused by the reduction of the black hole mass after the radiation of a small black hole.
The similar correction due to the back reaction in the process of radiation of particles was obtained by Parikh and Wilczek \cite{Wilczek2000} (see also Ref.\cite{Kraus1997}):
 \begin{equation}
p(\omega, M-\omega)\propto \exp{\left(-8\pi G\omega\left(M-\frac{\omega}{2}\right)\right)}
\,,
\label{tunnelingMomega}
\end{equation}
where $\omega$ is the energy of the emitted particle.
As distinct from Eq.(\ref{tunnelingMmquadratic}) for emission of a small black hole, the Eq.(\ref{tunnelingMomega}) contains the factor $1/2$.
The reason for such difference is that in case of the emission of a small black hole, the probability of emission contains the extra term compared to the emission of particles -- the entropy of the emitted black hole, $S(m)=4\pi G m^2$:
 \begin{equation}
p(m,M-m)\propto \exp{\left(-8\pi Gm\left(M-\frac{m}{2}\right)+ 4\pi G m^2\right)}
\,.
\label{tunnelingMm2}
\end{equation}
As a result the Eq.(\ref{tunnelingMmquadratic}) is restored. This also coincides with the result for the radiation of the self-gravitating shell.\cite{KrausWilczek1995}

Eq.(\ref{tunnelingM1M2}) can be extended to emission of several black holes. For example, the probability of emission of $N$ black holes with masses $m=M/N$ is:
\begin{eqnarray}
p(M \rightarrow mN) \propto \exp{\left(-4\pi G M^2\left(1 -\frac{1}{N}\right)\right)}\,.
\label{tunnelingMN}
\end{eqnarray}
For large $N$ this corresponds to the probability of the destruction of the black hole by explosion within the particular channel:
\begin{eqnarray}
p_{N\rightarrow \infty} \propto \exp(-S_{\rm BH})\,.
\label{tunnelingMNinf}
\end{eqnarray}
There are many channels of the destruction of the black hole in quantum tunnelling. The black hole entropy  $S_{\rm BH}$ can be estimated using the probability  of the black hole explosion in a single quantum event. Then, the Eq.(\ref{tunnelingMNinf}) suggests that the entropy of the  black hole is determined by the number of the possible channels leading to the destruction of the black hole.  

The discussed scenario of the destruction of black holes by explosion to the small Planck scale objects also explains the origin of the negative entropy of the white hole. The explosion is a rare process in which the entropy is reduced to zero. However, after the explosion occurs, the even the more rare process may follow, when the small Planck scale objects are collected back forming the white hole. The rarity of this process reduces the entropy even further, which leads to the negative entropy of the formed white hole. This is the reason why the white hole has the negative entropy. The latter is  also supported by the consideration of super-rare process of the quantum tunnelling from the black to the white hole of the same mass in Sec. \ref{WHentropy}. The decrease of the entropy in this super-rare process corresponds to the loss of the double entropy of the black hole, which results in the negative entropy of the white hole.

The negative temperature of the white hole is also not very surprising.  In general, the negative
absolute temperatures are consistent with equilibrium thermodynamics. All the thermodynamic properties,
such as thermometry, thermodynamics of cyclic transformations, ensemble equivalence, fluctuation-dissipation
relations, response theory, the transport processes and symmetry breaking phase transitions, can be reformulated to include the negative temperatures.\cite{Baldovin2021,Lounasmaa1997,Volovik2021}

\section{Conclusion}
\label{ConclusionSec}

The starting point of our consideration was that matter immersed in the de Sitter vacuum percieves this vacuum as the heat bath with the local temperature $T=H/\pi$, where $H$ is the Hubble parameter. This temperature has no relation to the cosmological horizon, and to the Hawking radiation from the cosmological horizon. However,  it is exactly twice the Gibbons-Hawking temperature, $T_{\rm GH}=H/2\pi$. The reason for such relation is the specific symmetry of the de Sitter space-time, which is similar to the invariance under translations in the Minkowski vacuum.

There are also the discrete symmetries, which are important for consideration of the thermodynamics of the de Sitter. The expanding de Sitter Universe represents one of the two degenerate states formed by the broken time reversal symmetry. These states are the expanding de Sitter state and the contracting de Sitter state. These states are obtained from each other by the time reversal transformation, $t\rightarrow -t$. Another broken discrete symmetry corresponds to the reversal of the sign of the scalar curvature, ${\cal R} \rightarrow -{\cal R}$. This symmetry operation transforms the de Sitter state to the anti-de Sitter. This symmetry is spontaneously broken by the term linear in ${\cal R}$ in the Einstein action, $\sqrt{-g} K{\cal R}$, where the gravitational coupling $K$ plays the role of the order parameter in this scenario of the symmetry breaking.

The local thermodynamics of the de Sitter state in the Einstein gravity gives rise to the Gibbons-Hawking area law for the total entropy inside the cosmological horizon. We extended the consideration of the local thermodynamics to the $f({\cal R})$ gravity and obtained the same area law, but with the modified gravitational coupling $K=df/d{\cal R}$. The agreement with the traditional global thermodynamics of de Sitter supports the suggestion that the de Sitter vacuum is the thermal state with the local temperature $T=H/\pi$, and that the local thermodynamics is based on the thermodynamically conjugate gravitational variables $K$ and ${\cal R}$. 
  This pair of the non-extensive gravitational variables is similar to the pair of the electrodynamic variables, electric field ${\bf E}$ and electric induction ${\bf D}$, which participate in the thermodynamics of dielectrics.
The gravitational variables modify the thermodynamic Gibbs-Duhem relation, due to which the thermal properties of the de Sitter state become similar to that of the Zel'dovich stiff matter and of the non-relativistic Fermi liquid.

The local temperature $T=H/\pi$ leads to the multiple creation of particles, if even only single electron is introduced to the de Sitter vacuum. This results in the thermal instability of the de Sitter state towards the formation of matter, and to further thermalization of this matter by the de Sitter heat bath. The process of thermalization of matter by the de Sitter heat bath, which takes place without the effects from the cosmological horizon, leads to the decay of the vacuum energy density.
As distinct from this process, the possibility of instability of the de Sitter state due to the Hawking radiation from the cosmological horizon is still controversial.

We considered two scenarios of the vacuum decay, which give two different power laws of decay.
One of them reproduces the result of the Padmanabhan model.\cite{Padmanabhan2005} The second one is based on the thermal fluctuations in the de Sitter heat bath. It leads to  the reasonable values of the dark energy and dark matter in the present time, and these values are not sensitive to the initial state of the Universe. This scenario suggests the simultaneous solution of three cosmological constant problems: why the cosmological constant is not large;  why the dark energy is on the order of dark matter; and why they have the present order of magnitude.

Using local thermodynamics with the local temperature $T=H/\pi$, we obtained the connection between the bulk entropy of the Hubble volume, and the surface entropy of the cosmological horizon $S_{\rm hor}=A/4G$. This suggests a kind of the bulk-surface  correspondence, which may have the holographic origin.\cite{Bekenstein1981,Verlinde2023,Milekhin2023} It would be interesting to check this correspondence using the more general extensions of the Einstein gravity and also different types of the generalized entropy.\cite{Odintsov2023,Odintsov2023a,Odintsov2022,Odintsov2021,Witten2023} It is important that such connection takes place only in the $3+1$ spacetime, where there is the special symmetry due to which both gravitational variables $K$ and ${\cal R}$ have the mass dimension 2, the same as the electrodynamic variables, electric field ${\bf E}$ and electric induction ${\bf D}$. It will be interesting to extend these considerations to another system with similar symmetry properties --  the Einstein static Universe, which also has constant curvature.

 We also discussed the thermodynamics of de Sitter in the frame of the statistical ensemble of the multi-metric gravities. The heat exchange between different "sub-Universes" in the ensemble leads to the common de Sitter expansion with the common temperature $T=H/\pi$.  Application of the local thermodynamics to the entropy of the Schwarzschild black hole was also considered. We obtained the Bekenstein-Hawking entropy of black hole from the negative entropy of the contracting de Sitter core of the gravastar object.
 



\end{document}